\documentclass[12pt,onecolumn]{mn2e}
\title[Polarization of WMAP point sources]
{High-frequency radio polarization measurements of WMAP point sources}
\author[N. Jackson et al.]
{N. Jackson$^{1}$, I.W.A. Browne$^{1}$, R.A. Battye$^{1}$, D.
Gabuzda$^{2}$, and A.C. Taylor$^{3}$
\\
$^{1}$Jodrell Bank Centre for Astrophysics, University of Manchester,
Turing Building, Oxford Road, Manchester M13 9PL\\
$^{2}$Physics Department, University College Cork, Ireland\\
$^{3}$Oxford Astrophysics, University of Oxford, Denys Wilkinson Building,
Keble Road, Oxford, OX1 3RH, UK\\
}

\def\PsfigVersion{1.9}
\ifx\undefined\psfig\else\endinput\fi

\let\LaTeXAtSign=\@
\let\@=\relax
\edef\psfigRestoreAt{\catcode`\@=\number\catcode`@\relax}
\catcode`\@=11\relax
\newwrite\@unused
\def\ps@typeout#1{{\let\protect\string\immediate\write\@unused{#1}}}
\ps@typeout{psfig/tex \PsfigVersion}

\def\figurepath{./}

\def\@nnil{\@nil}
\def\@empty{}
\def\@psdonoop#1\@@#2#3{}
\def\@psdo#1:=#2\do#3{\edef\@psdotmp{#2}\ifx\@psdotmp\@empty \else
    \expandafter\@psdoloop#2,\@nil,\@nil\@@#1{#3}\fi}
\def\@psdoloop#1,#2,#3\@@#4#5{\def#4{#1}\ifx #4\@nnil \else
       #5\def#4{#2}\ifx #4\@nnil \else#5\@ipsdoloop #3\@@#4{#5}\fi\fi}
\def\@ipsdoloop#1,#2\@@#3#4{\def#3{#1}\ifx #3\@nnil 
       \let\@nextwhile=\@psdonoop \else
      #4\relax\let\@nextwhile=\@ipsdoloop\fi\@nextwhile#2\@@#3{#4}}
\def\@tpsdo#1:=#2\do#3{\xdef\@psdotmp{#2}\ifx\@psdotmp\@empty \else
    \@tpsdoloop#2\@nil\@nil\@@#1{#3}\fi}
\def\@tpsdoloop#1#2\@@#3#4{\def#3{#1}\ifx #3\@nnil 
       \let\@nextwhile=\@psdonoop \else
      #4\relax\let\@nextwhile=\@tpsdoloop\fi\@nextwhile#2\@@#3{#4}}
\ifx\undefined\fbox
\newdimen\fboxrule
\newdimen\fboxsep
\newdimen\ps@tempdima
\newbox\ps@tempboxa
\long\def\fbox#1{\leavevmode\setbox\ps@tempboxa\hbox{#1}\ps@tempdima\fboxrule
    \advance\ps@tempdima \fboxsep \advance\ps@tempdima \dp\ps@tempboxa
   \hbox{\lower \ps@tempdima\hbox
  {\vbox{\hrule height \fboxrule
          \hbox{\vrule width \fboxrule \hskip\fboxsep
          \vbox{\vskip\fboxsep \box\ps@tempboxa\vskip\fboxsep}\hskip 
                 \fboxsep\vrule width \fboxrule}
                 \hrule height \fboxrule}}}}
\fi
\newread\ps@stream
\newif\ifnot@eof       
\newif\if@noisy        
\newif\if@atend        
\newif\if@psfile       
{\catcode`\%=12\global\gdef\epsf@start{
\def\epsf@PS{PS}
\def\epsf@getbb#1{%
\openin\ps@stream=#1
\ifeof\ps@stream\ps@typeout{Error, File #1 not found}\else
   {\not@eoftrue \chardef\other=12
    \def\do##1{\catcode`##1=\other}\dospecials \catcode`\ =10
    \loop
       \if@psfile
	  \read\ps@stream to \epsf@fileline
       \else{
	  \obeyspaces
          \read\ps@stream to \epsf@tmp\global\let\epsf@fileline\epsf@tmp}
       \fi
       \ifeof\ps@stream\not@eoffalse\else
       \if@psfile\else
       \expandafter\epsf@test\epsf@fileline:. \\%
       \fi
          \expandafter\epsf@aux\epsf@fileline:. \\%
       \fi
   \ifnot@eof\repeat
   }\closein\ps@stream\fi}%
\long\def\epsf@test#1#2#3:#4\\{\def\epsf@testit{#1#2}
			\ifx\epsf@testit\epsf@start\else
\ps@typeout{Warning! File does not start with `\epsf@start'.  It may not be a PostScript file.}
			\fi
			\@psfiletrue} 
{\catcode`\%=12\global\let\epsf@percent=
\long\def\epsf@aux#1#2:#3\\{\ifx#1\epsf@percent
   \def\epsf@testit{#2}\ifx\epsf@testit\epsf@bblit
	\@atendfalse
        \epsf@atend #3 . \\%
	\if@atend	
	   \if@verbose{
		\ps@typeout{psfig: found `(atend)'; continuing search}
	   }\fi
        \else
        \epsf@grab #3 . . . \\%
        \not@eoffalse
        \global\no@bbfalse
        \fi
   \fi\fi}%
\def\epsf@grab #1 #2 #3 #4 #5\\{%
   \global\def\epsf@llx{#1}\ifx\epsf@llx\empty
      \epsf@grab #2 #3 #4 #5 .\\\else
   \global\def\epsf@lly{#2}%
   \global\def\epsf@urx{#3}\global\def\epsf@ury{#4}\fi}%
\def\epsf@atendlit{(atend)} 
\def\epsf@atend #1 #2 #3\\{%
   \def\epsf@tmp{#1}\ifx\epsf@tmp\empty
      \epsf@atend #2 #3 .\\\else
   \ifx\epsf@tmp\epsf@atendlit\@atendtrue\fi\fi}

\chardef\psletter = 11 
\chardef\other = 12

\newif \ifdebug 
\newif\ifc@mpute 
\c@mputetrue 

\let\then = \relax
\def\r@dian{pt }
\let\r@dians = \r@dian
\let\dimensionless@nit = \r@dian
\let\dimensionless@nits = \dimensionless@nit
\def\internal@nit{sp }
\let\internal@nits = \internal@nit
\newif\ifstillc@nverging
\def \Mess@ge #1{\ifdebug \then \message {#1} \fi}

{ 
	\catcode `\@ = \psletter
	\gdef \nodimen {\expandafter \n@dimen \the \dimen}
	\gdef \term #1 #2 #3%
	       {\edef \t@ {\the #1}
		\edef \t@@ {\expandafter \n@dimen \the #2\r@dian}%
		\t@rm {\t@} {\t@@} {#3}%
	       }
	\gdef \t@rm #1 #2 #3%
	       {{%
		\count 0 = 0
		\dimen 0 = 1 \dimensionless@nit
		\dimen 2 = #2\relax
		\Mess@ge {Calculating term #1 of \nodimen 2}%
		\loop
		\ifnum	\count 0 < #1
		\then	\advance \count 0 by 1
			\Mess@ge {Iteration \the \count 0 \space}%
			\Multiply \dimen 0 by {\dimen 2}%
			\Mess@ge {After multiplication, term = \nodimen 0}%
			\Divide \dimen 0 by {\count 0}%
			\Mess@ge {After division, term = \nodimen 0}%
		\repeat
		\Mess@ge {Final value for term #1 of 
				\nodimen 2 \space is \nodimen 0}%
		\xdef \Term {#3 = \nodimen 0 \r@dians}%
		\aftergroup \Term
	       }}
	\catcode `\p = \other
	\catcode `\t = \other
	\gdef \n@dimen #1pt{#1} 
}

\def \Divide #1by #2{\divide #1 by #2} 

\def \Multiply #1by #2
       {{
	\count 0 = #1\relax
	\count 2 = #2\relax
	\count 4 = 65536
	\Mess@ge {Before scaling, count 0 = \the \count 0 \space and
			count 2 = \the \count 2}%
	\ifnum	\count 0 > 32767 
	\then	\divide \count 0 by 4
		\divide \count 4 by 4
	\else	\ifnum	\count 0 < -32767
		\then	\divide \count 0 by 4
			\divide \count 4 by 4
		\else
		\fi
	\fi
	\ifnum	\count 2 > 32767 
	\then	\divide \count 2 by 4
		\divide \count 4 by 4
	\else	\ifnum	\count 2 < -32767
		\then	\divide \count 2 by 4
			\divide \count 4 by 4
		\else
		\fi
	\fi
	\multiply \count 0 by \count 2
	\divide \count 0 by \count 4
	\xdef \product {#1 = \the \count 0 \internal@nits}%
	\aftergroup \product
       }}

\def\r@duce{\ifdim\dimen0 > 90\r@dian \then   
		\multiply\dimen0 by -1
		\advance\dimen0 by 180\r@dian
		\r@duce
	    \else \ifdim\dimen0 < -90\r@dian \then  
		\advance\dimen0 by 360\r@dian
		\r@duce
		\fi
	    \fi}

\def\Sine#1%
       {{%
	\dimen 0 = #1 \r@dian
	\r@duce
	\ifdim\dimen0 = -90\r@dian \then
	   \dimen4 = -1\r@dian
	   \c@mputefalse
	\fi
	\ifdim\dimen0 = 90\r@dian \then
	   \dimen4 = 1\r@dian
	   \c@mputefalse
	\fi
	\ifdim\dimen0 = 0\r@dian \then
	   \dimen4 = 0\r@dian
	   \c@mputefalse
	\fi
	\ifc@mpute \then
		\divide\dimen0 by 180
		\dimen0=3.141592654\dimen0
		\dimen 2 = 3.1415926535897963\r@dian 
		\divide\dimen 2 by 2 
		\Mess@ge {Sin: calculating Sin of \nodimen 0}%
		\count 0 = 1 
		\dimen 2 = 1 \r@dian 
		\dimen 4 = 0 \r@dian 
		\loop
			\ifnum	\dimen 2 = 0 
			\then	\stillc@nvergingfalse 
			\else	\stillc@nvergingtrue
			\fi
			\ifstillc@nverging 
			\then	\term {\count 0} {\dimen 0} {\dimen 2}%
				\advance \count 0 by 2
				\count 2 = \count 0
				\divide \count 2 by 2
				\ifodd	\count 2 
				\then	\advance \dimen 4 by \dimen 2
				\else	\advance \dimen 4 by -\dimen 2
				\fi
		\repeat
	\fi		
			\xdef \sine {\nodimen 4}%
       }}

\def\Cosine#1{\ifx\sine\UnDefined\edef\Savesine{\relax}\else
		             \edef\Savesine{\sine}\fi
	{\dimen0=#1\r@dian\advance\dimen0 by 90\r@dian
	 \Sine{\nodimen 0}
	 \xdef\cosine{\sine}
	 \xdef\sine{\Savesine}}}	      

\def\psdraft{
	\def\@psdraft{0}
}
\def\psfull{
	\def\@psdraft{100}
}

\psfull

\newif\if@scalefirst
\def\psscalefirst{\@scalefirsttrue}
\def\psrotatefirst{\@scalefirstfalse}
\psrotatefirst

\newif\if@draftbox
\def\psnodraftbox{
	\@draftboxfalse
}
\def\psdraftbox{
	\@draftboxtrue
}
\@draftboxtrue

\newif\if@prologfile
\newif\if@postlogfile
\def\pssilent{
	\@noisyfalse
}
\def\psnoisy{
	\@noisytrue
}
\psnoisy
\newif\if@bbllx
\newif\if@bblly
\newif\if@bburx
\newif\if@bbury
\newif\if@height
\newif\if@width
\newif\if@rheight
\newif\if@rwidth
\newif\if@angle
\newif\if@clip
\newif\if@verbose
\def\@p@@sclip#1{\@cliptrue}

\newif\if@decmpr

\def\@p@@sfigure#1{\def\@p@sfile{null}\def\@p@sbbfile{null}
	        \openin1=#1.bb
		\ifeof1\closein1
	        	\openin1=\figurepath#1.bb
			\ifeof1\closein1
			        \openin1=#1
				\ifeof1\closein1%
				       \openin1=\figurepath#1
					\ifeof1
					   \ps@typeout{Error, File #1 not found}
						\if@bbllx\if@bblly
				   		\if@bburx\if@bbury
			      				\def\@p@sfile{#1}%
			      				\def\@p@sbbfile{#1}%
							\@decmprfalse
				  	   	\fi\fi\fi\fi
					\else\closein1
				    		\def\@p@sfile{\figurepath#1}%
				    		\def\@p@sbbfile{\figurepath#1}%
						\@decmprfalse
	                       		\fi%
			 	\else\closein1%
					\def\@p@sfile{#1}
					\def\@p@sbbfile{#1}
					\@decmprfalse
			 	\fi
			\else
				\def\@p@sfile{\figurepath#1}
				\def\@p@sbbfile{\figurepath#1.bb}
				\@decmprtrue
			\fi
		\else
			\def\@p@sfile{#1}
			\def\@p@sbbfile{#1.bb}
			\@decmprtrue
		\fi}

\def\@p@@sfile#1{\@p@@sfigure{#1}}

\def\@p@@sbbllx#1{
		\@bbllxtrue
		\dimen100=#1
		\edef\@p@sbbllx{\number\dimen100}
}
\def\@p@@sbblly#1{
		\@bbllytrue
		\dimen100=#1
		\edef\@p@sbblly{\number\dimen100}
}
\def\@p@@sbburx#1{
		\@bburxtrue
		\dimen100=#1
		\edef\@p@sbburx{\number\dimen100}
}
\def\@p@@sbbury#1{
		\@bburytrue
		\dimen100=#1
		\edef\@p@sbbury{\number\dimen100}
}
\def\@p@@sheight#1{
		\@heighttrue
		\dimen100=#1
   		\edef\@p@sheight{\number\dimen100}
}
\def\@p@@swidth#1{
		\@widthtrue
		\dimen100=#1
		\edef\@p@swidth{\number\dimen100}
}
\def\@p@@srheight#1{
		\@rheighttrue
		\dimen100=#1
		\edef\@p@srheight{\number\dimen100}
}
\def\@p@@srwidth#1{
		\@rwidthtrue
		\dimen100=#1
		\edef\@p@srwidth{\number\dimen100}
}
\def\@p@@sangle#1{
		\@angletrue
		\edef\@p@sangle{#1} 
}
\def\@p@@ssilent#1{ 
		\@verbosefalse
}
\def\@p@@sprolog#1{\@prologfiletrue\def\@prologfileval{#1}}
\def\@p@@spostlog#1{\@postlogfiletrue\def\@postlogfileval{#1}}
\def\@cs@name#1{\csname #1\endcsname}
\def\@setparms#1=#2,{\@cs@name{@p@@s#1}{#2}}
\def\ps@init@parms{
		\@bbllxfalse \@bbllyfalse
		\@bburxfalse \@bburyfalse
		\@heightfalse \@widthfalse
		\@rheightfalse \@rwidthfalse
		\def\@p@sbbllx{}\def\@p@sbblly{}
		\def\@p@sbburx{}\def\@p@sbbury{}
		\def\@p@sheight{}\def\@p@swidth{}
		\def\@p@srheight{}\def\@p@srwidth{}
		\def\@p@sangle{0}
		\def\@p@sfile{} \def\@p@sbbfile{}
		\def\@p@scost{10}
		\def\@sc{}
		\@prologfilefalse
		\@postlogfilefalse
		\@clipfalse
		\if@noisy
			\@verbosetrue
		\else
			\@verbosefalse
		\fi
}
\def\parse@ps@parms#1{
	 	\@psdo\@psfiga:=#1\do
		   {\expandafter\@setparms\@psfiga,}}
\newif\ifno@bb
\def\bb@missing{
	\if@verbose{
		\ps@typeout{psfig: searching \@p@sbbfile \space  for bounding box}
	}\fi
	\no@bbtrue
	\epsf@getbb{\@p@sbbfile}
        \ifno@bb \else \bb@cull\epsf@llx\epsf@lly\epsf@urx\epsf@ury\fi
}	
\def\bb@cull#1#2#3#4{
	\dimen100=#1 bp\edef\@p@sbbllx{\number\dimen100}
	\dimen100=#2 bp\edef\@p@sbblly{\number\dimen100}
	\dimen100=#3 bp\edef\@p@sbburx{\number\dimen100}
	\dimen100=#4 bp\edef\@p@sbbury{\number\dimen100}
	\no@bbfalse
}
\newdimen\p@intvaluex
\newdimen\p@intvaluey
\def\rotate@#1#2{{\dimen0=#1 sp\dimen1=#2 sp
		  \global\p@intvaluex=\cosine\dimen0
		  \dimen3=\sine\dimen1
		  \global\advance\p@intvaluex by -\dimen3
		  \global\p@intvaluey=\sine\dimen0
		  \dimen3=\cosine\dimen1
		  \global\advance\p@intvaluey by \dimen3
		  }}
\def\compute@bb{
		\no@bbfalse
		\if@bbllx \else \no@bbtrue \fi
		\if@bblly \else \no@bbtrue \fi
		\if@bburx \else \no@bbtrue \fi
		\if@bbury \else \no@bbtrue \fi
		\ifno@bb \bb@missing \fi
		\ifno@bb \ps@typeout{FATAL ERROR: no bb supplied or found}
			\no-bb-error
		\fi
		\count203=\@p@sbburx
		\count204=\@p@sbbury
		\advance\count203 by -\@p@sbbllx
		\advance\count204 by -\@p@sbblly
		\edef\ps@bbw{\number\count203}
		\edef\ps@bbh{\number\count204}
		\if@angle 
			\Sine{\@p@sangle}\Cosine{\@p@sangle}
	        	{\dimen100=\maxdimen\xdef\r@p@sbbllx{\number\dimen100}
					    \xdef\r@p@sbblly{\number\dimen100}
			                    \xdef\r@p@sbburx{-\number\dimen100}
					    \xdef\r@p@sbbury{-\number\dimen100}}
                        \def\minmaxtest{
			   \ifnum\number\p@intvaluex<\r@p@sbbllx
			      \xdef\r@p@sbbllx{\number\p@intvaluex}\fi
			   \ifnum\number\p@intvaluex>\r@p@sbburx
			      \xdef\r@p@sbburx{\number\p@intvaluex}\fi
			   \ifnum\number\p@intvaluey<\r@p@sbblly
			      \xdef\r@p@sbblly{\number\p@intvaluey}\fi
			   \ifnum\number\p@intvaluey>\r@p@sbbury
			      \xdef\r@p@sbbury{\number\p@intvaluey}\fi
			   }
			\rotate@{\@p@sbbllx}{\@p@sbblly}
			\minmaxtest
			\rotate@{\@p@sbbllx}{\@p@sbbury}
			\minmaxtest
			\rotate@{\@p@sbburx}{\@p@sbblly}
			\minmaxtest
			\rotate@{\@p@sbburx}{\@p@sbbury}
			\minmaxtest
			\edef\@p@sbbllx{\r@p@sbbllx}\edef\@p@sbblly{\r@p@sbblly}
			\edef\@p@sbburx{\r@p@sbburx}\edef\@p@sbbury{\r@p@sbbury}
		\fi
		\count203=\@p@sbburx
		\count204=\@p@sbbury
		\advance\count203 by -\@p@sbbllx
		\advance\count204 by -\@p@sbblly
		\edef\@bbw{\number\count203}
		\edef\@bbh{\number\count204}
}
\def\in@hundreds#1#2#3{\count240=#2 \count241=#3
		     \count100=\count240	
		     \divide\count100 by \count241
		     \count101=\count100
		     \multiply\count101 by \count241
		     \advance\count240 by -\count101
		     \multiply\count240 by 10
		     \count101=\count240	
		     \divide\count101 by \count241
		     \count102=\count101
		     \multiply\count102 by \count241
		     \advance\count240 by -\count102
		     \multiply\count240 by 10
		     \count102=\count240	
		     \divide\count102 by \count241
		     \count200=#1\count205=0
		     \count201=\count200
			\multiply\count201 by \count100
		 	\advance\count205 by \count201
		     \count201=\count200
			\divide\count201 by 10
			\multiply\count201 by \count101
			\advance\count205 by \count201
		     \count201=\count200
			\divide\count201 by 100
			\multiply\count201 by \count102
			\advance\count205 by \count201
		     \edef\@result{\number\count205}
}
\def\compute@wfromh{
		\in@hundreds{\@p@sheight}{\@bbw}{\@bbh}
		\edef\@p@swidth{\@result}
}
\def\compute@hfromw{
	        \in@hundreds{\@p@swidth}{\@bbh}{\@bbw}
		\edef\@p@sheight{\@result}
}
\def\compute@handw{
		\if@height 
			\if@width
			\else
				\compute@wfromh
			\fi
		\else 
			\if@width
				\compute@hfromw
			\else
				\edef\@p@sheight{\@bbh}
				\edef\@p@swidth{\@bbw}
			\fi
		\fi
}
\def\compute@resv{
		\if@rheight \else \edef\@p@srheight{\@p@sheight} \fi
		\if@rwidth \else \edef\@p@srwidth{\@p@swidth} \fi
}
\def\compute@sizes{
	\compute@bb
	\if@scalefirst\if@angle
	\if@width
	   \in@hundreds{\@p@swidth}{\@bbw}{\ps@bbw}
	   \edef\@p@swidth{\@result}
	\fi
	\if@height
	   \in@hundreds{\@p@sheight}{\@bbh}{\ps@bbh}
	   \edef\@p@sheight{\@result}
	\fi
	\fi\fi
	\compute@handw
	\compute@resv}

\def\psfig#1{\vbox {
	\ps@init@parms
	\parse@ps@parms{#1}
	\compute@sizes
	\ifnum\@p@scost<\@psdraft{
		\special{ps::[begin] 	\@p@swidth \space \@p@sheight \space
				\@p@sbbllx \space \@p@sbblly \space
				\@p@sbburx \space \@p@sbbury \space
				startTexFig \space }
		\if@angle
			\special {ps:: \@p@sangle \space rotate \space} 
		\fi
		\if@clip{
			\if@verbose{
				\ps@typeout{(clip)}
			}\fi
			\special{ps:: doclip \space }
		}\fi
		\if@prologfile
		    \special{ps: plotfile \@prologfileval \space } \fi
		\if@decmpr{
			\if@verbose{
				\ps@typeout{psfig: including \@p@sfile.Z \space }
			}\fi
			\special{ps: plotfile "`zcat \@p@sfile.Z" \space }
		}\else{
			\if@verbose{
				\ps@typeout{psfig: including \@p@sfile \space }
			}\fi
			\special{ps: plotfile \@p@sfile \space }
		}\fi
		\if@postlogfile
		    \special{ps: plotfile \@postlogfileval \space } \fi
		\special{ps::[end] endTexFig \space }
		\vbox to \@p@srheight sp{
			\hbox to \@p@srwidth sp{
				\hss
			}
		\vss
		}
	}\else{
		\if@draftbox{		
			\hbox{\frame{\vbox to \@p@srheight sp{
			\vss
			\hbox to \@p@srwidth sp{ \hss \@p@sfile \hss }
			\vss
			}}}
		}\else{
			\vbox to \@p@srheight sp{
			\vss
			\hbox to \@p@srwidth sp{\hss}
			\vss
			}
		}\fi

	}\fi
}}
\psfigRestoreAt
\let\@=\LaTeXAtSign

\begin{document}
\maketitle
\begin{abstract}

We present polarization measurements at 8.4, 22, and 43~GHz made with
the VLA of a complete sample of extragalactic sources stronger than 1~Jy in the
5-year WMAP catalogue and with declinations north of $-34^{\circ}$.  The
observations were motivated by the need to know the polarization
properties of radio sources at frequencies of tens of GHz in order to
subtract polarized foregrounds for future sensitive Cosmic Microwave
Background (CMB) experiments. The total intensity and polarization 
measurements are generally consistent with comparable VLA calibration 
measurements for less-variable sources, and within a similar
range to WMAP fluxes for unresolved sources. A further paper will
present correlations between measured parameters and derive implications
for CMB measurements.
\end{abstract}


\large
\baselineskip 15pt

\section{Introduction}

Radio source samples, selected at high frequency and with flux densities
of the order of 1~Jy, mostly contain compact
flat spectrum objects associated with the nuclei of active galaxies
which have relativistic outflows pointing close to the direction of
the observer (e.g.  Blandford \& Rees 1978; Scheuer \& Readhead 1979;
Orr \& Browne 1982). This emission consists of a number of synchrotron
components, each with a steep optically thin synchrotron spectrum at
high frequencies and a self-absorbed spectrum at lower frequencies;
the combination of many such components with different cutoff
frequencies produces the observed ``flat'' radio spectrum. These
components are frequently linearly polarized, and in principle useful
information on jet physics can be obtained from the distribution of
polarization in frequency and space. Although high frequency-selected
samples are dominated by flat spectrum objects the samples do contain
a minority of intrinsically more powerful sources in which the
extended, steep-spectrum synchrotron emission dominates rather than
the emission from a beamed core. Such sources also display linearly
polarized emission, often at a higher level than the flat spectrum
sources.

In recent years there has been particular interest in high-frequency
properties of radio sources because discrete radio sources act as a
confusing foreground for measurements of the Cosmic Microwave
Background (e.g. Tegmark \& Efstathiou 1996). Such sources contribute
most to the power spectrum on small angular scales (high $\ell$), and
efforts have been made to subtract their effects in several different
ways. For total intensity, high frequency observations of sources
selected from lower frequency surveys have been used. For example, the
Very Small Array CMB interferometer, working at 31~GHz, used 
15-GHz source-finding observations
with the Ryle Telescope followed by monitoring of discrete sources
with dedicated long baselines (Dickinson et al. 2004) and/or single
dish observations made with the Torun 32~m (Gawronski et al. 2009),
while for the Cosmic Background Imager experiment high frequency
observations were made with the OVRO 42~m and GBT of sources selected
at 1.4~GHz (Mason et al. 2003; Mason et al. 2009). Alternatively,
population models based on extrapolations from lower frequency
catalogues can be used statistically to ameliorate the effects of the
extragalactic source foreground.

Although large-scale blind surveys at high frequencies are difficult
to do, because of the small field of view at high frequencies, there
are a number of both blind and pointed surveys in existence. The
southern hemisphere has been completely surveyed by the AT20G survey
(Ricci et al. 2004), which has been conducted with the Australia
Telescope Compact Array (ATCA) at 18~GHz down to a limiting flux
density of 50~mJy. A detailed study of brighter sources in this survey
has been presented by Massardi et al. (2008). The whole sky has been
covered to a brighter limiting flux density of about 1Jy by WMAP at a
number of frequencies $\geq$20~GHz (Wright et al. 2009), and smaller
areas of the northern sky have been covered to greater depth by
surveys such as the 15-GHz 9C survey (Waldram et al. 2003). These
surveys generally find that sources that are bright at high
frequencies have complex spectra. About 30\% are significantly
inverted (Ricci et al. 2004), and the mean spectral index around
20--40 GHz tends to be around zero, generally steepening at higher
frequencies (Ricci et al. 2006; Massardi et al. 2008; Vol'Vach et
al. 2008). There is some evidence that the spectra of most sources 
turn over by 95~GHz (Sadler et al. 2008) but in general extrapolation 
to both higher frequencies and to lower flux density levels is needed 
in order to realistically simulate the CMB discrete foregrounds (e.g. 
Waldram et al. 2007). Importantly, there is convincing evidence that the
mix of flat and steep spectrum sources changes with flux density. This
is a prediction of beaming models (e.g. Orr \& Browne 1982; Wall \&
Jackson 1997). There is also direct observational evidence provided by
Gawronski et al. (2009) that the proportion of steep-spectrum sources
with extended radio emission increases from $\leq$20\% in a WMAP 22-GHz
sample to $\geq$50\% in Ryle-selected 15-GHz sample which has a flux
density limit about two orders of magnitude weaker than that of the WMAP
sample. Thus care must be taken when extrapolating to both higher
frequencies and lower flux densities. The sensitivity of instruments 
such as Planck require this to be done as accurately as possible, and
hence it is important to base the models on as firm an observational
footing as possible.

The situation with polarized foregrounds is less developed but it is
becoming urgent to solve because inflation models make different
predictions for the strength of B-modes in the CMB polarization
distribution. A heroic effort is going into experiments to
detect B-modes. These include Planck and ground-based projects like
QUIET (Samtleben 2008) and the balloon-borne experiment SPIDER (Montroy 
et al. 2006). All CMB observations are technically challenging
because they rely on extreme stability of the equipment and on
meticulous subtraction of foregrounds to get at the true distribution
of the CMB fluctuations. The detection of the B-mode signal will be
particularly difficult (though just how difficult depends on
its strength which varies over a wide range depending on the inflation
model). For scalar-to-tensor ratio r=0.01, the peak signal (at around
1$^{\circ}$ scales) is $\sim$30nK.  B-mode experiments operate at high
frequencies (e.g. QUIET: 44 and 90GHz; Spider: 90, 145 and 220~GHz;
PLANCK 30,70, 100, 143, 217 and 353~GHz). Hence observations at the
highest frequencies possible are desirable. Estimates of synchrotron
source total intensity contamination have been made by extrapolating
from observations made at 1.4 and 5~GHz (e.g. Tucci et al. 2004). The
current best-guess models are presented by Toffolatti et
al. (1998, 2005). However, additional assumptions need to be made if the
polarized foreground is to be characterized; one needs to know about
the high frequency polarization properties of sources. Ideally one
would like direct observations of polarized source counts but this is
impractical. An alternative is to use high frequency total intensity
source counts and a knowledge of how the percentage polarizations of
sources change with both frequency and flux density to predict
polarized source counts.



The work reported in this paper is the first phase in a programme
aimed at improving knowledge about the polarization properties of
radio sources at high radio frequencies. We begin with bright sources
from the WMAP sample of Wright et al. (2009) which have $S_{\rm
22GHz}>$1~Jy.  In this work, all but four of the sample of 203 WMAP
sources with flux density of $\geq$1~Jy at 22~GHz and with declination
$\delta>-34$ have been imaged with the VLA in total intensity and
polarization. The second phase of our programme will be to measure the
polarization properties of a high frequency-selected sample of much
weaker sources designed to look for any dependence of polarization
properties on flux density. Given that Gawronski et al. (2009) find that
the population mix changes with flux density there is every reason to
expect that average polarization properties also change.

\section{Observations and data reduction}

Observations were conducted using the VLA in D-configuration between
01.40 and 19.30 UT on 2008 August 1. During this period 16
antennas were equipped with EVLA receivers while the remaining 11 had
original VLA receivers. Each source was normally observed at three
frequencies: two 50-MHz bands centred on 8.4351 and 8.4851
GHz (X-band), two 50-MHz bands centred on 22.4351 and 22.4851 GHz
(K-band), and two 50-MHz bands centred on 43.3149 and 43.3649
MHz (Q-band). Pointing calibration was carried out approximately every 
15-20 minutes. The available time allowed observations to be made 
of all the sources at
22~GHz and 43~GHz, and approximately two-thirds of the sources at
8~GHz. Four sources (WMAP~0823+224, WMAP~0824+392, WMAP~1014+231 and
WMAP~1840+797) were missed during the observations, leaving a total of
199 sources out of 203 with attempted measurements. The source 3C48
was observed in order to provide an additional flux and polarization
calibrator; the primary flux and polarization calibrator, 3C286
(WMAP~1331+305) was already part of the sample. Additionally, three
much weaker sources from the 9C survey were also observed as a pilot
for a  future observing programme aimed at much weaker sources.

The nominal time on each source was 37 seconds at 8~GHz, 30 seconds
at 22~GHz and 80 seconds at 43~GHz, giving a theoretical rms of 0.2,
0.5 and 0.7 mJy/beam, respectively, for natural weighting. However,
the rms noise in most images is typically greater than this, being
between 1~mJy/beam and 2~mJy/beam in the majority of images. This is
partly due to the use of uniform weighting during the imaging process,
which increases the noise level typically by factors between 1.2 and
1.5.  However, in particular, 8-GHz observations were affected by a
combination of missing antennas and slightly longer than predicted
slew and setup times on many sources. A sampling time of 3 seconds per
integration was used, the minimum allowed by the observing system.

All data analysis and processing was carried out in the {\sc aips}
package, distributed by the US National Radio Astronomy Observatory. 
Data were examined and flagged manually. The 22-GHz data were relatively
clean, but extensive flagging of bad telescopes was performed source by
source on both 8-GHz and 43-GHz data, resulting in the removal of up to
half of the data on any given source. The major problems were loss of
correlation on some telescopes during setup at the beginning of a scan,
together with intermittent bad data on many individual telescopes.

Overall amplitude calibration was performed using observations of the
source 3C286 at each frequency and normalising to the flux scale of
Baars et al. (1977). Amplitude and phase solutions were made using the
sources stronger than a limiting flux density; the process was iterated,
adjusting this limiting flux, until the amplitude and phase solutions
appeared to vary in a coherent manner. Amplitude solutions were smoothed
using an averaging time of 1 hour for 43~GHz and 0.1 hours for the other
frequencies. In the case of the 8-GHz observations, a further
calibration was made, again using a standard model for 3C286, for 
baseline-dependent offsets. This was necessary because of the mixture 
of receivers in the array, approximately equally distributed between old
(VLA) and new (EVLA) receivers.

Polarization calibration was then carried out. Telescope polarization
offsets were calibrated by use of the source 3C84, which was assumed
to have zero polarization. This assumption is correct for the level of
accuracy required here: Taylor et al. (2006) report levels of $<$0.1\%
polarization in the core of the source at frequencies less than 22 GHz, 
with approximately 0.2\% at 22GHz, and a somewhat higher level
(0.8-7.5\%) in the weak mas-scale jet component. The overall level of
polarization in 3C84 is likely to be about 0.1\%. 3C84 was not observed 
at 8~GHz. In this case the source 0125$-$001, a bright point source for 
which no significant polarization was detected in either 22-GHz or 43-GHz
observations, was used instead. The polarization position angle was
calibrated by use of the 3C286 data, applying corrections to the R$-$L
phase difference to rotate the 3C286 polarization position angle to
33$^{\circ}$. The baseline-to-baseline scatter in this quantity was
0.6, 3.3 and 8.5 degrees for 8~GHz, 22~GHz and 43~GHz respectively.

Imaging was performed using the AIPS task IMAGR with uniform
weighting, on a default 128$\times$128 grid and cellsizes of 1\farcs8,
0\farcs7 and 0\farcs35 for 8~GHz, 22~GHz and 43~GHz 
respectively. For strong sources (roughly speaking,
those containing sufficient flux density to have been used as phase
calibrators), one iteration of phase-only self-calibration and one
iteration of amplitude self-calibration was applied. Final total
intensity images were produced using 1000 {\sc clean} iterations
(H\"ogbom 1974); images in Stokes $Q$ and $U$ were produced using a
limited number (40) of iterations to avoid reduction in the signal due
to {\sc clean} bias (e.g. Condon et al. 1998).

Estimation of the Stokes $I$, $Q$ and $U$ flux densities was performed
in two ways. The first method consisted of fitting to the calibrated 
($u,v$) data, constraining the fit by fixing the position to the peak of
the $I$ map, constraining the intrinsic size of the peak component to be
$<1^{\prime\prime}$, but allowing the $Q$ and $U$ flux density to vary. 
Constraints are necessary to avoid instability in the fits due to 
noise in the polarization data. The alternative method consisted of
finding, and fitting to, the highest absolute value in the $Q$ and $U$
images, within a square of side 10 pixels. Neither method is ideal.
Fitting to the ($u,v$) data has the advantage of independence from effects
due to deconvolution, notably the effects of {\sc clean} bias. On the
other hand, stabilising the fit requires fixing the peak of the Stokes
$Q$ and $U$ to the peak of the $I$ data, and this may not be appropriate
for sources with extended polarized structure. Fig. 1 shows a comparison
of the two methods, and demonstrates that for the majority of sources
the two methods agree well. Approximately 10\% of sources show
significant disagreement, mostly those with low polarized flux
densities. In Table 2, where the data are presented, detailed 
comments are given in cases where the two methods disagree. Errors given
in Table 2 are a quadrature combination of the random error, as output
by {\sc aips}, and the systematic calibration error, assumed to be 5\%
for 43 GHz and 22~GHz and 2\% at 8.4~GHz.

\begin{figure*}
\psfig{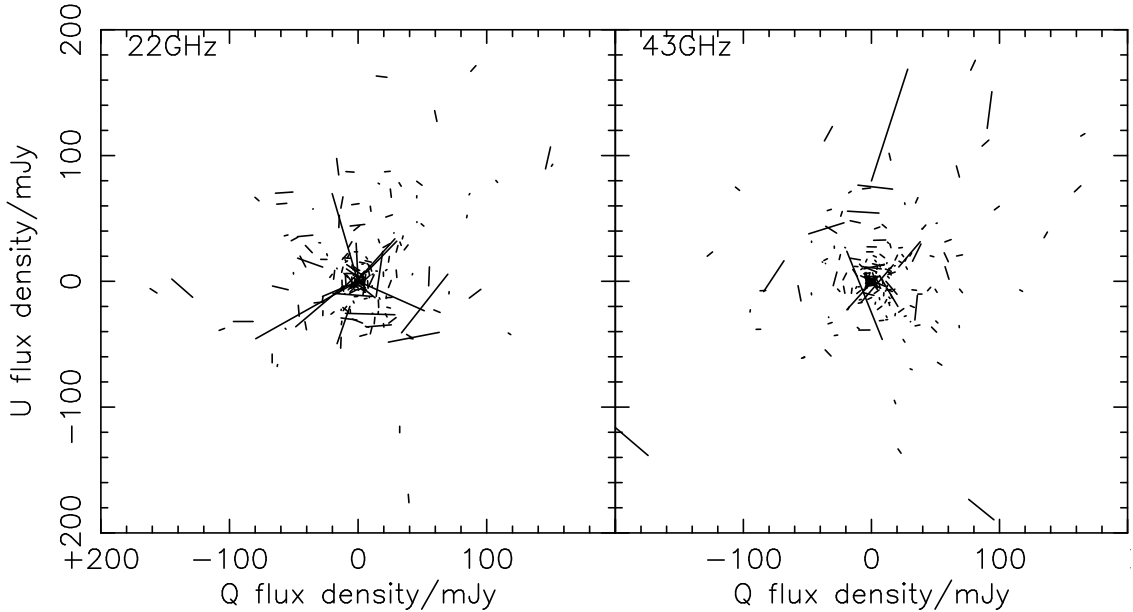}
\caption{Estimates of the Stokes parameters $Q$ and $U$ for the mapped
sources, in mJy, for the two methods described in the text (fitting to
the images and to the ($u,v$) data). In each case, $Q$, in mJy, is the 
abscissa and $U$, in mJy, is the ordinate, and lines on the plot join
the ($Q,U$) coordinates of the first measurement to those of the second.
Note that for most sources the methods agree 
well and the lines are relatively short. In cases of
disagreement we provide a note in Table 2 explaining the choice of
polarized flux densities used in the data tables.}
\end{figure*}

Because the total polarization is a quadratic sum of $Q$ and $U$, it is
positive-biased; values of polarized flux density of three times the rms
noise cannot therefore be regarded as significant detections. Many
studies of this effect have been made. For example, Simmons \& Stewart
(1985) give a relation of estimated signal-to-noise, $p_o$, and 
observed signal-to-noise $p$, in polarization for four different indicators.
These all converge towards equality for $p>4$ as $p_o=(p^2-1)^{1/2}$,
and diverge strongly for $p<2$, reaching values of $1<p<1.5$ for
$p_o=0$. The issue in these observations is complicated by the different
values of rms noise in different images. Inspection of the images
suggests that polarized flux densities of 10mJy are likely to be 
significant at all frequencies in both methods.

Fig. 2 shows the complete total intensity and Stokes maps, together with
the spectrum in total intensity and polarization, for the first source
in order of right ascension, WMAP~0006$-$063. The other images can be
found in the online material associated with this article.

\begin{figure*}
\psfig{figure=0006-063.ps,width=9cm,angle=-90}
\psfig{figure=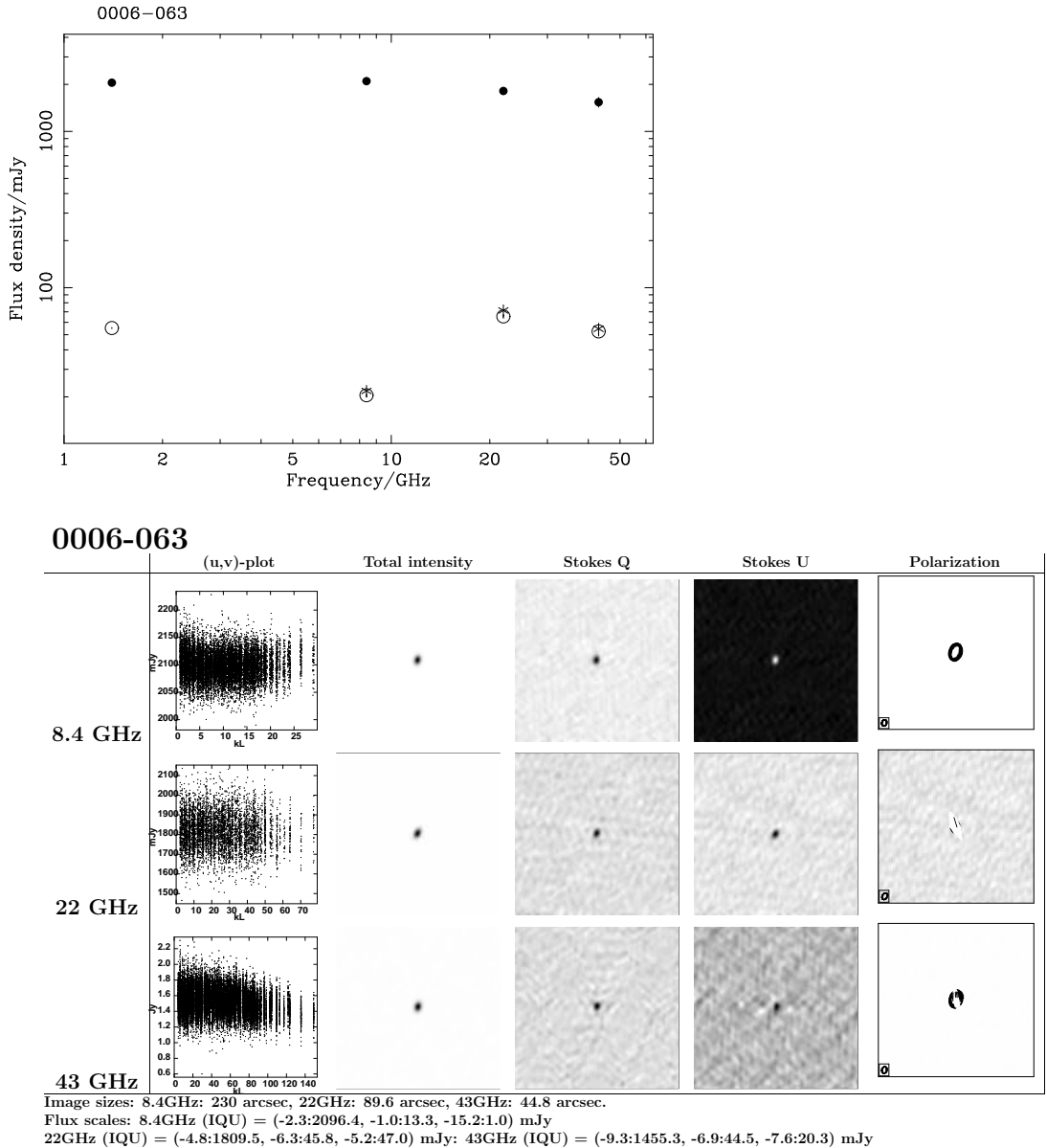,width=15cm,angle=90}
\caption{Full data for the first WMAP source, 0006$-$063. The spectrum
(top) is given in units of log (flux density/mJy) against log
(frequency/GHz); error bars are shown but  
are smaller than the plot symbols in many cases.
Filled circles are total intensity points, and open
circles are polarized flux density points measured by fitting to the ($u,v$) data.
Stars are polarized flux density points measured by maxima from the $Q,U$ maps.
The maps of the source (bottom) include, in columns from left to right,
a plot of the correlated amplitude as a function of baseline length,
maps in Stokes $I$, $Q$ and $U$, and a plot of the polarized flux
vectors superimposed upon contours of Stokes $I$. Rows, from top to
bottom, correspond to maps at 8, 22 and 43~GHz.}
\end{figure*}

\begin{table*}
\begin{tabular}{lccccccc} \hline
Frequency&1.4&&8.4&&22&&43\\
(GHz) & &1.4-8.4&&8.4-22&&22-43&\\ \hline
Polarization (\%) & 2.2$\pm$0.2 && 3.1$\pm$0.2 && 2.7$\pm$0.2 && 3.1$\pm$0.2\\
                  &   (1.6)     &&    (2.6)    &&    (2.2)    &&   (2.6) \\
Intensity spectral index && 0.01$\pm$0.04 && $-$0.20$\pm$0.05 && $-$0.49$\pm$0.04 & \\
                         &&   (0.04)      &&      ($-$0.15)   && ($-$0.37) & \\
Polarization spectral index && 0.35$\pm$0.07 && 0.01$\pm$0.10 && $-$0.09$\pm$0.06 & \\ 
               &&  (0.28) && ($-$0.15) && ($-$0.14)& \\ \hline
\end{tabular}
\caption{Average intensity and polarization quantities for the sample,
counting polarization limits as zero. 1.4-GHz observations are from
NVSS, 8.4-GHz observations from CLASS and this work, and the remainder
from this work. Median values are given in brackets. Spectral indices
are defined as $S_{\nu}=\nu^{\alpha}$.}
\end{table*}

\section{Comparison with other observations}

\subsection{Consistency of polarization with the VLA calibrator programme}

A few polarization calibrators are observed regularly by the VLA (Taylor
\& Myers 2000)\footnote{Online data from the VLA polarization
calibration programme is available on
http://www.vla.nrao.edu/astro/calib/polar/.}. We compare the total
intensity and polarization measurements of these sources with our 
measurements for 22~GHz and 43~GHz in Fig. 3. Sources included in this
comparison are those which were observed by the VLA in the
D-configuration of 2008, interpolated if necessary to the date of our
observation. Those sources which appear non-variable to 15\% are plotted
as separate symbols; it appears that our measurements compare well with
the VLA calibration programme. 

\begin{figure*}
\psfig{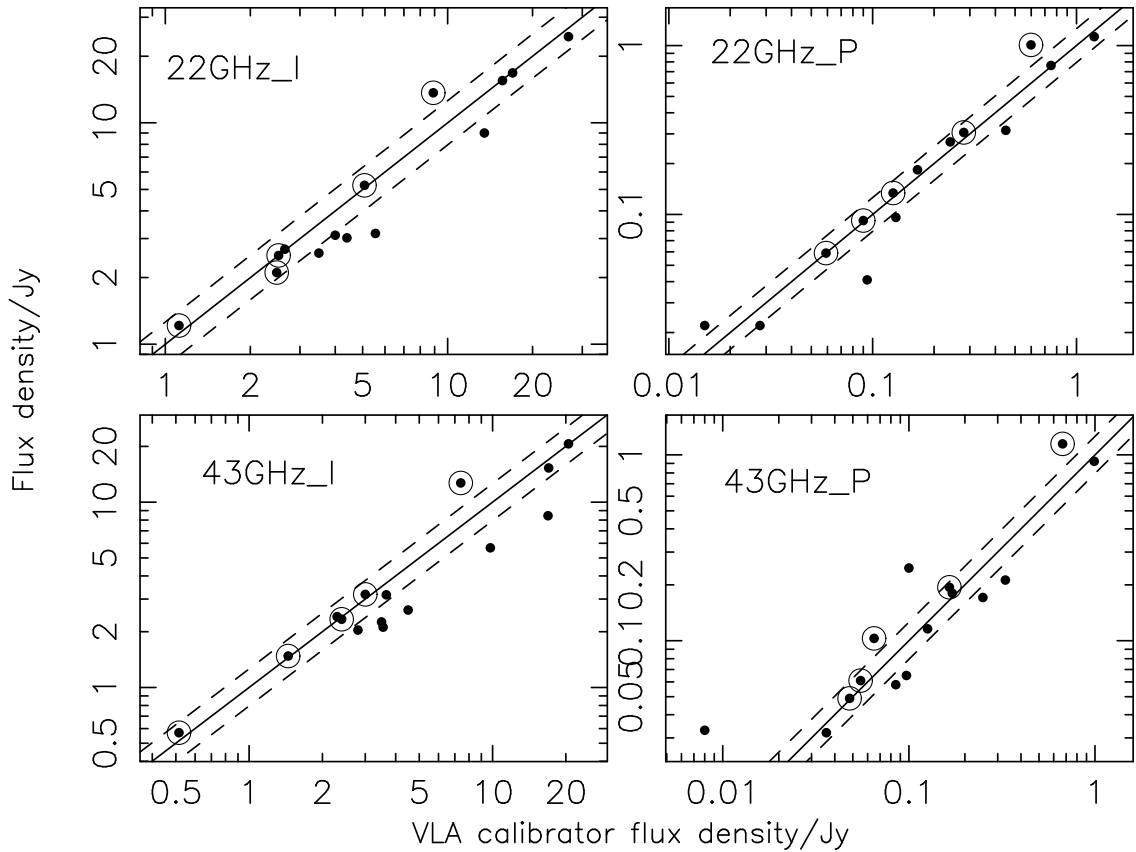}
\caption{Comparison of intensity and polarized flux density measurements
between this work and the VLA polarization calibrator monitoring
programme (Taylor \& Myers 2000). Sources are plotted if they were
observed in the D-configuration of 2008 by the monitoring programme,
interpolated to the date of our observation if necessary.
Encircled points represent sources which were stable during the period
3C286, 3C48, 1310+323, 1924$-$292, 2136+006). Dashed lines represent
variation of 20\% from the line of equal flux density between these
observations and VLA calibration monitoring observations.}
\end{figure*}

\subsection{Comparison with WMAP fluxes}

Ricci et al. (2006) comment that their AT20G flux densities from ATCA data on
southern sources are systematically different from those of WMAP, in
that the WMAP 30~GHz observations of flat spectrum sources seem to be
low by a factor of about 1.2. Differences in this sense cannot be the
result of resolution effects, since the larger beam of WMAP should
result in a higher detected flux density in any resolved sources. We
can investigate this with the northern-hemisphere 22-GHz data presented
here, and the results are shown in Fig. 4.

\begin{figure}
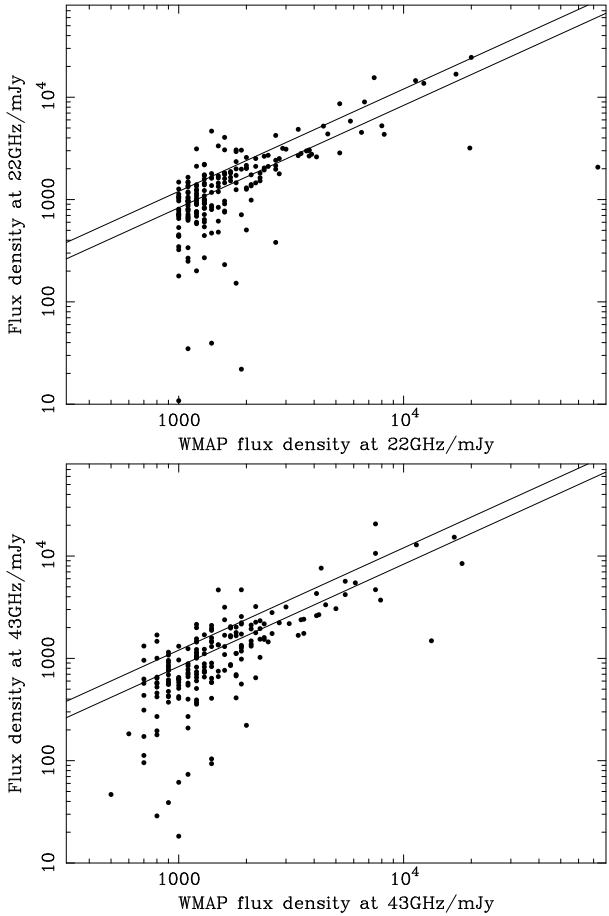

\begin{tabular}{c}
\psfig{figure=K_vla_wmap.ps,width=8cm,angle=-90}\\
\psfig{figure=Q_vla_wmap.ps,width=8cm,angle=-90}\\
\end{tabular}
\caption{Comparison of the 22-GHz and 43-GHz flux densities in this work
with those of Wright et al. (2009; quoted to two significant figures in
that paper). Lines are drawn at 20\% either side
of the line of equal flux.}
\end{figure}

It is clear that there is a population of resolved sources, for which
the WMAP flux densities at 22 and 43~GHz are higher than those of this
work. A population of sources unresolved with either set of observations
cluster around the line of equal flux density, although there will
be some scatter induced by variability of sources between the two epochs
of observation. At the 20\% level, there is no evidence for the flux
scales being significantly different at 22~GHz. There is marginal
evidence at 43~GHz for slightly lower flux densities in these
observations than in the WMAP observations, which may be an indication
that even some relatively compact sources are beginning to be resolved
at this frequency. It may also be a consequence of variability bias, in
which some sources which are on average just below the 1-Jy limit, 
happened to be in a relatively high state when observed with WMAP. More
generally, the WMAP flux densities are significantly affected by CMB 
fluctuations, since a $\sim 50\mu$K fluctuation in a 30$^{\prime}$ beam 
corresponds to a flux density of about 100~mJy between 20--30~GHz.

\subsection{Comparison with CLASS}

We now compare the 8-GHz data from this project with the
polarization measurements made in CLASS (Jackson et al. 2008), typically
between 1990 and 1992, which are
shown in Fig. 5 for sources with significant (6$\sigma$) polarization
detections in both works. As expected, the total flux densities cluster
around equal values for these point sources, but the polarized flux
densities are lower for CLASS. Again, this is likely to be a resolution
effect, since the CLASS observations were taken in A-configuration of
the VLA, which has a maximum baseline a factor of 35 greater than that
of the D-configuration used for these observations. We pick up more 
polarized flux in these D-configuration observations. The implication is
that there is polarized emission present on arcsecond scales and 
therefore resolved out by CLASS. The
relative degrees of resolution of the total intensity and polarized flux
indicates that the structure being resolved out is more highly
polarized than that which is not.

\begin{figure}
\psfig{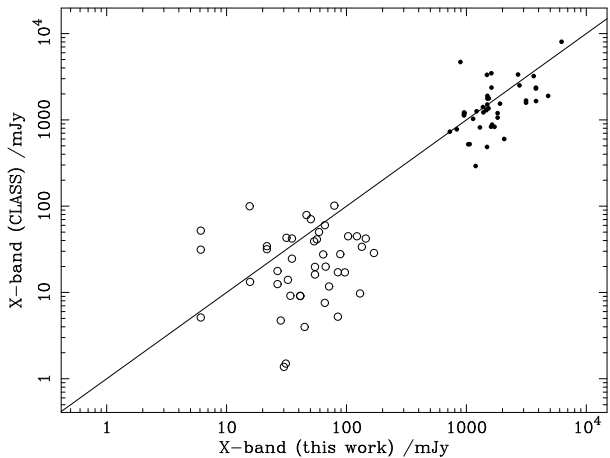}
\caption{Comparison between flux densities in this work and CLASS of
total flux density (filled circles) and polarized flux density (open
circles) for cases where both surveys detect significant polarized flux.
Note the generally lower polarized flux density in CLASS.}
\end{figure}

\subsection{Position angle differences and multiple components}

In Fig. 6 we plot the histograms of differences in position angle
between 8-GHz and 22-GHz polarization, and between 22-GHz and 43-GHz
polarization, for objects with significant detections of polarization,
have contemporaneous polarization measurements at all three frequencies,
and have no obviously heavily resolved structure. It is obvious that the two
lower frequencies are less well correlated in polarization position
angle. Correlation is, unsurprisingly, slightly worse if the CLASS
8-GHz observations are used in addition to the contemporaneous 8-GHz
observations. There is no obvious difference in position angle
correlation between brighter and fainter sources.

\begin{figure}
\psfig{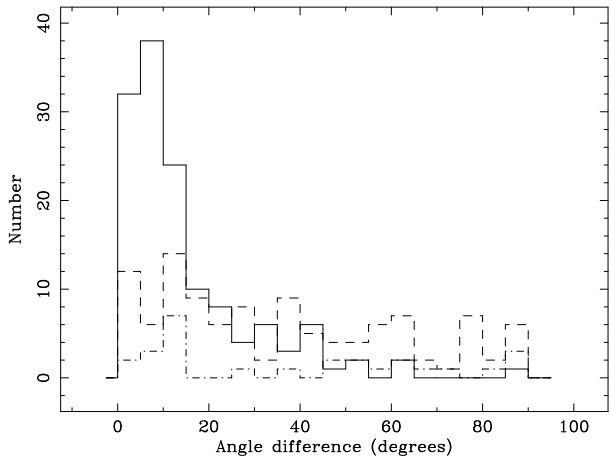}
\caption{Differences in position angle between observed polarization at
different frequencies. Solid line: difference between 22~GHz and 43~GHz.
Dashed line: difference between 8 and 22~GHz for contemporaneous
observations. Dash-dotted line: difference between 8 and 22~GHz for
objects where only archival CLASS 8-GHz data is available. Only
observations with significant detections of polarized flux (defined as
6$\sigma$ in CLASS and as objects with visible polarized flux in the $Q$
and $U$ maps in these observations) are plotted. Known heavily resolved
sources are excluded.}
\end{figure}

The obvious interpretation of these data are that significant Faraday
rotation is present at the lower frequencies; since this effect is
proportional to $\lambda^2$, the angle of rotation would be expected 
to be about 6-7 times larger between the lower two
frequencies. Galactic Faraday rotation is unlikely to be responsible for
this effect, since the $<$100~rad$\,$m$^{-2}$ typical of the Galactic
foreground is an order of magnitude smaller than the rotations needed
significantly to affect 8-GHz measurement, and indeed in our data there
is no correlation of this rotation with galactic latitude. Zavala \& 
Taylor (2004), in a study of parsec-scale polarization of quasar radio 
cores, find core RMs from 500 to a few thousand rad~m$^{-2}$, and suggest
that these are due to a foreground Faraday screen close to the radio jet.
Such parsec-scale components are likely to dominate the 22- and 43-GHz 
flux densities. On the larger scales probed by the lower-frequency 
observations, it is possible that in many sources there are multiple polarized
components with different position angles. The fact that more polarization
is detected here in D-configuration than in the A-configuration CLASS 
observations suggests that a more extended jet component, with a lower 
RM than the core, may be present. Future high-sensitivity observations 
with e-MERLIN and the EVLA should reveal any such low surface-brightness
components.

\section{Conclusions}

We have presented observations at 22 and 43~GHz of 199 of the 203 radio 
sources in the WMAP5 list of Wright et al. (2009) which have flux density
$>$1Jy at 22~GHz, and observations at 8.4~GHz of 133 of these objects.
The total flux densities are broadly consistent with those listed by
WMAP. For the few objects with previous VLA observations of polarized
flux densities, there is also good agreement. There is evidence for
extra polarized flux density on arcsecond scales at 8.4~GHz compared to earlier
observations at higher resolution. Polarization position angles are very
similar at 22 and 43~GHz, but are much less well correlated with
position angles measured at 8.4~GHz. This is likely to be due to
multiple polarized components which make up the generally complex radio
spectra of flat-spectrum radio sources. In a further paper we discuss
correlations between the measured intensity and polarization parameters
in more detail, and calculate the implications for future CMB
polarization measurements of the distribution of polarized flux
densities in these discrete sources.

\section*{Online material}

The complete maps can be accessed at 
{\tt http://www.jb.man.ac.uk/$\sim$njj/pol\_maps/}.
This figure includes, for each source, in columns from left to right, a 
plot of the correlated amplitude as a function of baseline length for
each source, maps in Stokes $I$, $Q$ and $U$, and a plot of the
polarized flux vectors superimposed upon contours of Stokes $I$. For
each object, rows are given corresponding to maps at 8, 22 and 43~GHz
respectively. 

Plots of spectra can be accessed at
{\tt http://www.jb.man.ac.uk/$\sim$njj/pol\_spec/}
and are given in units of log(flux density/mJy) against
log(frequency/GHz). Filled circles are total intensity points, and open
circles are polarized flux points measured by fitting to the ($u,v$) data. 
Stars are polarized flux points measured by maxima from the $Q$, $U$ maps.

\section*{Acknowledgements}

The Very Large Array, National Radio Astronomy Observatory, is a 
facility of the National Science Foundation operated under cooperative 
agreement by Associated Universities, Inc.

\section*{References}

\noindent Baars J.W.M., Genzel R., Pauliny-Toth I.I.K., Witzel A. 1977,  A\&A, 61, 99. 

\noindent Blandford R.D., Rees M.J. 1978, Phys S, 17, 265

\noindent Condon J.J., Cotton W.D., Greisen E.W., Yin Q.F., Perley R.A., Taylor G.B., Broderick J.J. 1998,  AJ, 115, 1693. 

\noindent Dickinson C., Battye R.A., Carreira P., Cleary K., Davies R.D., Davis R.J., Genova-Santos R.,
Grainge K., Guti\'errez C.M., Hafez Y.A.,  2004,  MNRAS, 353, 732. 

\noindent Gawronski M.P., et al., 2009, in preparation

\noindent Gizani N.A.B., Leahy J.P. 2003,  MNRAS, 342, 399. 

\noindent Hogbom J.A. ,  A\&AS, 15, 417. 

\noindent Laing R.A. 1981,  MNRAS, 195, 261. 

\noindent Leahy J.P., Perley R.A. 1991,  AJ, 102, 537. 

\noindent Leahy J.P., Black A.R.S., Dennett-Thorpe J., Hardcastle M.J., Komissarov S., Perley R.A., Riley J.M., Scheuer P.A.G. 1997,  MNRAS, 291, 20. 

\noindent Mason B.S., Pearson T.J., Readhead A.C.S., Shepherd M.C., Sievers J., Udomprasert P.S., Cartwright J.K., Farmer A.J., Padin S., Myers S.T.,  2003,  ApJ, 591, 540. 

\noindent Mason B.S., Weintraub L.C., Sievers J.L., Bond J.R., Myers
S.T., Perason T.J., Readhead A.C.S., Shepherd M.C., 2009, astro-ph/0901.4330

\noindent Massardi M., Ekers R.D., Murphy T., Ricci R., Sadler E.M., Burke S.,
deZotti G., Edwards P.G., Hancock P.J., Jackson C.A.,  2008,  MNRAS, 384, 775. 

\noindent Montroy, T.E., et al. 2006, Ground-based and Airborne Telescopes. Edited
by Stepp, L.M., Proc. SPIE, 6267, 62670R

\noindent Orr M.J.L., Browne I.W.A. 1982,  MNRAS, 200, 1067. 

\noindent Perley R.A., Bridle A.H., Willis A.G. 1984,  ApJS, 54, 291. 

\noindent Reid A., Shone D.L., Akujor C.E., Browne I.W.A., Murphy D.W., Pedelty J., Rudnick L., Walsh D. ,  A\&AS, 110, 213. 

\noindent Ricci R., Prandoni I., Gruppioni C., Sault R.J., DeZotti G. 2004,  A\&A, 415, 549. 

\noindent Ricci R., Prandoni I., Gruppioni C., Sault R.J., deZotti G. 2006,  A\&A, 445, 465. 

\noindent Sadler E.M., Ricci R., Ekers R.D.., Ekers J.A., Hancock P.J., Jackson C.A., Kesteven M.J., Murphy T.,
Phillips C., Reinfrank R.F.,  2006,  MNRAS, 371, 898. 

\noindent Sadler E.M., Ricci R., Ekers R.D.., Sault RobertJ., Jackson C.A., deZotti G. 2008,  MNRAS, 385, 1656. 

\noindent Samtleben D., 2008, To be published in the Proceedings of the
43rd "Rencontres de Moriond" on Cosmology 2008, arXiv0806.4334

\noindent Saikia D.J., Kulkarni V.K., Porcas R.W. 1986,  MNRAS, 219, 719. 

\noindent Saikia D.J., Subrahmanya C.R., Patnaik A.R., Unger S.W., Cornwell T.J., Graham D.A., Prabhu T.P. 1986,  MNRAS, 219, 545. 

\noindent Scheuer P.A.G., Readhead A.C.S. 1979,  Natur, 277, 182. 

\noindent Simmons J.F.L., Stewart B.G. 1985,  A\&A, 142, 100. 

\noindent Taylor G.B., Gugliucci N.E., Fabian A.C., Sanders J.S.,
Gentile G., Allen S.W., 2006, MNRAS 368, 1500

\noindent Taylor G.B., Myers S.T., 2000, VLA Scientific Memo 26.
Available from NRAO or from http://www.vlba.nrao.edu/memos/sci/

\noindent Tegmark M., Efstathiou G. 1996,  MNRAS, 281, 1297. 

\noindent Toffolatti L., Argueso Gomez F., de Zotti G., Mazzei P, Franceschini A., Danese L., Burigana
C., 1998, MNRAS 297, 116

\noindent Toffolatti L., Negrello M., Gonz\'alez-Nuevo J., de Zotti G., Silva
L., Granato G.L., Argueso F., 2005, A\&A 438, 475

\noindent Tucci M., Mart\'{\i}nez-Gonz\'alez E., Toffolatti L.,
Gonz\'alez-Nuevo J., de Zotti G., 2004, MNRAS 349, 1267

\noindent Vol'Vach A.E., Vol'Vach L.N., Kardashev N.S., Larionov M.G. 2008,  ARep 52, 429. 

\noindent Waldram E.M., Pooley G.G., Grainge K.J.B., Jones M.E., Saunders R.D.E., Scott P.F., Taylor A.C. 2003,  MNRAS 342, 915. 

\noindent Waldram E.M., Bolton R.C., Pooley G.G., Riley J.M. 2007,  MNRAS 379, 1442. 

\noindent Wall J.V., Jackson C.A., 1997, MNRAS 290, L17

\noindent Wright E.L., Chen X., Odegard N., Bennett C.L., Hill R.S., Hinshaw G., Jarosik N., Komatsu E., Nolta M.R., Page L.,  2009,  ApJS 180, 283. 

\noindent Zavala R.T., Taylor G.B. 2004, ApJ 612, 749

\begin{table*}
\begin{tabular}{lcccccccccc}
&\multicolumn{3}{c}{X-band}&\multicolumn{3}{c}{K-band}&\multicolumn{3}{c}{Q-band}\\
&I(mJy)&P(mJy)&Angle($^{\circ}$)&I(mJy)&P(mJy)&Angle($^{\circ}$)&I(mJy)&P(mJy)&Angle($^{\circ}$)\\ \hline
0006$-$063 &2100$\pm$40 & 20.5$\pm$0.4 & 155.9$\pm$0.5 & 1820$\pm$90 & 65.4$\pm$2.5 & 22.8$\pm$1.1 & 1540$\pm$80 & 52.5$\pm$2.5 & 10.9$\pm$0.9 & \\
0010+110 &{\em 469.0$\pm$0.1} & {\em 0.2$\pm$0.1} & {\em 87$\pm$7} & 710$\pm$40 & 6.3$\pm$2.4$\dagger$ & 115$\pm$10 & 1550$\pm$80 & 13.4$\pm$1.3 & 29$\pm$3 & \\
0019+203 &{\em 1275.1$\pm$0.4} & {\em 55.8$\pm$0.2} & {\em 106.9$\pm$0.1} & 650$\pm$30 & 7.1$\pm$1.9$\dagger$ & 149$\pm$12 & 475$\pm$24 &  $<$10 &  - & \\
0019+260 &389$\pm$8 & 8.1$\pm$0.2 & 9.4$\pm$0.7 & 347$\pm$17 &  $<$10 &  - & 270$\pm$14 &  $<$10 &  - & 1\\
0029+059 &740$\pm$15 & 19.8$\pm$0.3 & 116.0$\pm$0.5 & 630$\pm$30 & 17.5$\pm$2.0 & 130$\pm$4 & 423$\pm$21 & 15.9$\pm$1.3 & 146.9$\pm$2.6 & \\
0043+521 & & & & 179$\pm$11 &  $<$10 &  - & 46.7$\pm$2.9 &  $<$10 &  - & 2\\
0047$-$252 & & & & 268$\pm$16 &  $<$10 &  - & 73$\pm$8 &  $<$10 &  - & 3\\
0050$-$068 &940$\pm$19 & 23.0$\pm$0.5 & 144.8$\pm$0.4 & 1370$\pm$70 & 36$\pm$3 & 145$\pm$5 & 1320$\pm$70 & 39.1$\pm$2.1 & 127.9$\pm$1.2 & \\
0051$-$094 & & & & 1480$\pm$70 & 62.6$\pm$2.8 & 156.6$\pm$1.3 & 1470$\pm$70 & 77$\pm$3 & 147.2$\pm$0.9 & 4\\
0108+015 &1620$\pm$30 & 66.0$\pm$1.0 & 112.9$\pm$0.5 & 1980$\pm$100 & 38.2$\pm$2.4 & 103.3$\pm$1.3 & 1670$\pm$80 & 29.1$\pm$2.0 & 115.3$\pm$2.2 & \\
0108+133 &900$\pm$27 & 78.4$\pm$1.8 & 95.8$\pm$0.3 & 39.5$\pm$2.3 &  $<$10 &  - & 28.9$\pm$1.9 &  $<$10 &  - & 5\\
0116$-$116 &912$\pm$18 & 19.0$\pm$0.4 & 105.6$\pm$0.5 & 910$\pm$50 &  $<$10 &  - & 960$\pm$50 & 13.1$\pm$1.3 & 171.1$\pm$2.2 & \\
0121+118 &3810$\pm$80 & 59.0$\pm$1.2 & 90.3$\pm$0.1 & 3130$\pm$160 & 178$\pm$7 & 15.8$\pm$0.9 & 1990$\pm$100 & 143$\pm$7 & 8.0$\pm$0.6 & \\
0125$-$001 &1089$\pm$22 &  $<$10 &  - & 1030$\pm$50 &  $<$10 &  - & 780$\pm$40 & 4.7$\pm$2.1$\dagger$ & 111$\pm$14 & 6\\
0132$-$168 &830$\pm$17 & 18.0$\pm$0.4 & 104.5$\pm$0.6 & 1250$\pm$60 & 13$\pm$6$\dagger$ & 11$\pm$6 & 1660$\pm$80 & 39.4$\pm$2.1 & 164.1$\pm$1.4 & 7\\
0137$-$244 &949$\pm$19 & 19.1$\pm$0.4 & 113.3$\pm$0.6 & 910$\pm$50 &  $<$10 &  - & 690$\pm$30 & 16.8$\pm$2.0 & 12$\pm$3 & \\
0137+478 &3140$\pm$60 & 26.6$\pm$0.6 & 116.3$\pm$0.5 & 2660$\pm$130 & 59.1$\pm$3.0 & 71.5$\pm$1.3 & 1750$\pm$90 & 19.1$\pm$1.4 & 58.5$\pm$2.3 & \\
0149+058 &1137$\pm$23 & 25.4$\pm$0.4 & 112.7$\pm$0.5 & 730$\pm$40 &  $<$10 &  - & 421$\pm$21 &  $<$10 &  - & \\
0152+221 &{\em 1038.5$\pm$0.1} & {\em 20.9$\pm$0.1} & {\em 56.8$\pm$0.1} & 890$\pm$40 & 42.4$\pm$2.4 & 3.5$\pm$0.8 & 730$\pm$40 & 46.5$\pm$2.5 & 177.1$\pm$0.9 & 8\\
0204+152 &1485$\pm$30 & 30.1$\pm$0.6 & 142.1$\pm$0.3 & 820$\pm$40 & 6.3$\pm$1.0$\dagger$ & 58$\pm$5 & 547$\pm$27 & 18.8$\pm$1.3$\dagger$ & 168.9$\pm$1.9 & \\
0205+322 &{\em 626.8$\pm$2.3} & {\em 5.2$\pm$0.1} & {\em 48.4$\pm$0.4} & 3060$\pm$150 & 102$\pm$6 & 99.1$\pm$0.7 & 2150$\pm$110 & 87$\pm$4 & 92.4$\pm$0.6 & \\
0218+016 &{\em 1209.5$\pm$0.7} & {\em 17.7$\pm$0.3} & {\em 149.7$\pm$0.4} & 1720$\pm$90 & 20.1$\pm$1.6 & 5.7$\pm$1.9 & 1690$\pm$80 & 31.8$\pm$1.5 & 19.0$\pm$1.3 & \\
0220+359 &1720$\pm$30 & 134.2$\pm$2.4 & 35.8$\pm$0.2 & 1400$\pm$70 & 194$\pm$8 & 31.0$\pm$0.8 & 950$\pm$50 & 145$\pm$5 & 25.4$\pm$1.0 & \\
0223+430 &231$\pm$6 & 9.5$\pm$0.5 & 170.3$\pm$1.1 & 152$\pm$8 &  $<$10 &  - & 94$\pm$6 &  $<$10 &  - & 9\\
0231+133 &{\em 1732.2$\pm$1.4} & {\em 11.5$\pm$0.2} & {\em 157.0$\pm$0.6} & 1390$\pm$70 & 9.4$\pm$1.3$\dagger$ & 157$\pm$4 & 1030$\pm$50 & 11.1$\pm$1.8$\dagger$ & 117$\pm$4 & \\
0237+288 &3640$\pm$70 & 144.8$\pm$2.1 & 67.6$\pm$0.4 & 3050$\pm$150 & 49.3$\pm$2.6 & 78.0$\pm$1.2 & 2380$\pm$120 & 47.9$\pm$1.9 & 112.4$\pm$1.2 & \\
0238+166 &3190$\pm$60 & 30.8$\pm$0.7 & 91.7$\pm$0.3 & 3340$\pm$170 & 85$\pm$4 & 34.0$\pm$0.8 & 3160$\pm$160 & 57.7$\pm$2.3 & 23.2$\pm$1.1 & \\
0241$-$083 & & & & 1060$\pm$50 &  $<$10 &  - & 960$\pm$50 & 11.3$\pm$1.2 & 140.0$\pm$2.7 & 10\\
0259$-$002 &1067$\pm$21 & 25.5$\pm$0.4 & 116.5$\pm$0.5 & 860$\pm$40 & 14.0$\pm$2.4 & 55$\pm$3 & 508$\pm$25 &  $<$10 &  - & \\
0308+040 &833$\pm$21 & 29.3$\pm$0.8 & 105.0$\pm$0.6 & 542$\pm$27 & 6.4$\pm$1.1$\dagger$ & 63$\pm$5 & 372$\pm$19 & 6.1$\pm$1.1$\dagger$ & 116$\pm$5 & 11\\
0309+104 &{\em 1243.1$\pm$1.1} & {\em 25.9$\pm$0.3} & {\em 124.5$\pm$0.3} & 810$\pm$40 & 33.9$\pm$2.1 & 74.5$\pm$1.6 & 760$\pm$40 & 34.1$\pm$1.6 & 63.9$\pm$1.4 & \\
0319+415 & & & & 14530$\pm$730 & 12$\pm$3 & 99$\pm$7 & 10620$\pm$530 & 51$\pm$5 & 57.3$\pm$1.8 & 12\\
0329$-$239 &1137$\pm$23 & 35.3$\pm$0.6 & 157.0$\pm$0.5 & 1250$\pm$60 & 13.2$\pm$1.5 & 104$\pm$4 & 1040$\pm$50 & 11.2$\pm$1.9 & 110$\pm$5 & \\
0336$-$129 & & & & 453$\pm$23 & 11.4$\pm$1.7 & 171$\pm$7 & 399$\pm$20 & 13.5$\pm$1.3 & 163.6$\pm$2.7 & \\
0339$-$017 & & & & 2040$\pm$100 & 67$\pm$4 & 91.6$\pm$0.4 & 1780$\pm$90 & 48.0$\pm$2.5 & 96.5$\pm$1.1 & \\
0340$-$213 &1054$\pm$21 & 26.9$\pm$0.6 & 132.3$\pm$0.4 & 900$\pm$50 & 14$\pm$3 & 39$\pm$4 & 720$\pm$40 & 14.6$\pm$1.4 & 30.3$\pm$2.9 & \\
0348$-$277 & & & & 760$\pm$40 & 46$\pm$4 & 41.8$\pm$1.5 & 760$\pm$40 & 41.8$\pm$2.6 & 43$\pm$5 & \\
0358+104 &75$\pm$4 & 18.2$\pm$0.5 & 103.4$\pm$0.8 & 6.7$\pm$1.1 &  $<$10 &  - & 6.4$\pm$1.1 &  $<$10 &  - & 13\\
0405$-$130 &2230$\pm$40 & 31.1$\pm$0.6 & 157.9$\pm$0.5 & 1300$\pm$70 & 68$\pm$4 & 169.4$\pm$1.0 & 850$\pm$40 & 22.4$\pm$2.3 & 166.4$\pm$2.3 & 14\\
0411+769 &2060$\pm$40 & 48.9$\pm$0.8 & 120.4$\pm$0.4 & 1060$\pm$50 & 71$\pm$4 & 176.7$\pm$0.9 & 570$\pm$29 & 44.2$\pm$1.9 & 156.6$\pm$1.2 & \\
0416$-$208 &1103$\pm$23 & 40.7$\pm$0.8 & 121.8$\pm$0.4 & 650$\pm$30 & 11.8$\pm$2.1$\dagger$ & 52$\pm$5 & 544$\pm$27 &  $<$10 &  - & \\
0423$-$013 &3920$\pm$80 & 157.9$\pm$3.0 & 97.6$\pm$0.2 & 4340$\pm$220 & 145$\pm$8 & 89.6$\pm$0.3 & 3710$\pm$190 & 130$\pm$6 & 85.6$\pm$0.6 & \\
0423+023 &266$\pm$5 & 5.6$\pm$0.3 & 126.0$\pm$2.4 & 201$\pm$10 & 7.3$\pm$0.9 & 124$\pm$4 & 113$\pm$6 & 7.2$\pm$1.3 & 116$\pm$5 & 15\\
0424+005 &518$\pm$10 & 21.7$\pm$0.5 & 88.0$\pm$0.4 & 481$\pm$24 & 20.0$\pm$1.8 & 60.4$\pm$2.2 & 411$\pm$21 & 20.4$\pm$1.6 & 43.0$\pm$1.9 & \\
0433+053 &3660$\pm$70 & 90.1$\pm$1.7 & 144.0$\pm$0.3 & 2650$\pm$130 & 126$\pm$6 & 170.2$\pm$0.7 & 1950$\pm$100 & 86$\pm$3 & 154.7$\pm$1.0 & \\
0453$-$281 &1960$\pm$40 & 43$\pm$8 & 123.8$\pm$2.4 & 1700$\pm$90 & 21$\pm$5 & 63$\pm$5 & 1440$\pm$70 & 14.1$\pm$2.5 & 11$\pm$6 & \\
0456$-$233 &2080$\pm$40 & 44.7$\pm$1.1 & 120.3$\pm$0.6 & 1950$\pm$100 & 11.7$\pm$2.3$\dagger$ & 117$\pm$5 & 1530$\pm$80 & 18.4$\pm$2.7$\dagger$ & 106$\pm$4 & 16\\
0501$-$019 &995$\pm$20 & 8.3$\pm$0.3 & 109.7$\pm$0.9 & 960$\pm$50 & 6.3$\pm$1.1 & 33$\pm$6 & 890$\pm$40 & 33.6$\pm$1.8 & 34.1$\pm$1.0 & \\
0513$-$219 &1041$\pm$21 & 7.5$\pm$0.6 & 122.1$\pm$1.7 & 800$\pm$40 & 43.7$\pm$2.8 & 136$\pm$4 & 620$\pm$30 & 38.7$\pm$2.6 & 134.1$\pm$1.2 & \\
0519$-$056 &8.1$\pm$0.3 &  $<$10 &  - & 3.8$\pm$1.4 &  $<$10 &  - & 2.6$\pm$0.9 &  $<$10 &  - & 17\\
0527$-$126 & & & & 470$\pm$50 &  $<$10 &  - & 104$\pm$13 &  $<$10 &  - & 18\\
0542+498 &4680$\pm$90 & 119.9$\pm$2.1 & 168.9$\pm$0.3 & 1820$\pm$90 & 67.8$\pm$2.8 & 73.9$\pm$1.0 & 820$\pm$40 & 41.4$\pm$2.1 & 79.1$\pm$1.0 & \\
0555+397 &{\em 7200$\pm$18} & {\em 51.8$\pm$0.1} & {\em 64.4$\pm$0.1} & 3110$\pm$160 & 21.6$\pm$1.7 & 158.4$\pm$2.3 & 2040$\pm$100 & 32.9$\pm$1.7 & 151.9$\pm$1.4 & 19\\
0607+673 &{\em 611.5$\pm$0.4} & {\em 21.5$\pm$0.3} & {\em 121.8$\pm$0.3} & 600$\pm$30 &  $<$10 &  - & 435$\pm$23 & 49$\pm$18$\dagger$ & 20$\pm$11 & \\
0608$-$223 &1376$\pm$28 & 28.8$\pm$0.7 & 89.2$\pm$0.4 & 1470$\pm$70 & 68$\pm$4 & 50.0$\pm$2.2 & 1150$\pm$60 & 74$\pm$4 & 45.3$\pm$0.8 & \\
0609$-$156 & & & & 2990$\pm$150 & 54$\pm$5 & 52.8$\pm$1.4 & 2190$\pm$110 & 40$\pm$6 & 68$\pm$4 & \\
0629$-$199 &682$\pm$14 & 8.8$\pm$0.3 & 53.4$\pm$0.9 & 610$\pm$30 & 7.9$\pm$1.7$\dagger$ & 27$\pm$6 & 407$\pm$20 & 9.2$\pm$1.9$\dagger$ & 173$\pm$13 & \\
0636$-$205 &283$\pm$8 & 42.0$\pm$1.0 & 116.6$\pm$0.7 & 34.9$\pm$2.9 &  $<$10 &  - & 9.2$\pm$2.1 &  $<$10 &  - & 20\\
0646+448 &3800$\pm$80 & 84.7$\pm$1.7 & 44.8$\pm$0.1 & 3160$\pm$160 & 21.8$\pm$1.3 & 122.9$\pm$1.5 & 2120$\pm$110 & 31.9$\pm$2.0 & 140.6$\pm$1.8 & \\
0721+713 &2070$\pm$40 & 67.1$\pm$1.4 & 180.0$\pm$0.1 & 4050$\pm$200 & 91$\pm$12 & 28$\pm$6 & 4680$\pm$230 & 127$\pm$6 & 52.0$\pm$0.5 & \\
0738+177 &891$\pm$18 & 15.6$\pm$0.4 & 147.4$\pm$0.8 & 640$\pm$30 & 21.0$\pm$1.9 & 160.7$\pm$2.3 & 407$\pm$20 & 14.9$\pm$1.9 & 155$\pm$4 & \\
0739+016 &{\em 1809.3$\pm$0.9} & {\em 58.9$\pm$0.3} & {\em 45.1$\pm$0.1} & 1520$\pm$80 & 45.1$\pm$2.3 & 22.0$\pm$1.5 & 1480$\pm$70 & 29.6$\pm$1.8 & 9.4$\pm$1.2 & \\
\end{tabular}
\end{table*}

\clearpage
\begin{table*}
\begin{tabular}{lcccccccccc}
&\multicolumn{3}{c}{X-band}&\multicolumn{3}{c}{K-band}&\multicolumn{3}{c}{Q-band}\\
&I(mJy)&P(mJy)&Angle($^{\circ}$)&I(mJy)&P(mJy)&Angle($^{\circ}$)&I(mJy)&P(mJy)&Angle($^{\circ}$)\\ \hline
0741+311 &1610$\pm$30 & 79.2$\pm$1.5 & 36.9$\pm$0.2 & 760$\pm$40 & 19.1$\pm$3.0 & 74$\pm$7 & 458$\pm$23 & 17.8$\pm$1.3 & 68.9$\pm$2.0 & \\
0745+102 & & & & 970$\pm$50 & 10.3$\pm$1.5$\dagger$ & 14$\pm$3 & 312$\pm$16 &  $<$10 &  - & 21\\
0746$-$007 &1910$\pm$40 & 65.7$\pm$1.3 & 93.7$\pm$0.2 & 1060$\pm$50 & 27.9$\pm$1.5 & 147.1$\pm$1.3 & 567$\pm$28 & 18.4$\pm$1.5 & 139$\pm$7 & \\
0750+125 &4810$\pm$110 & 129.7$\pm$1.9 & 69.4$\pm$0.4 & 4230$\pm$210 & 148$\pm$7 & 33.2$\pm$0.8 & 2800$\pm$140 & 113$\pm$4 & 27.1$\pm$1.0 & \\
0753+539 &967$\pm$19 & 103.2$\pm$1.8 & 34.2$\pm$0.3 & 1020$\pm$50 & 91$\pm$4 & 37.0$\pm$0.7 & 910$\pm$50 & 76$\pm$3 & 30.7$\pm$1.0 & \\
0757+099 &1118$\pm$22 & 47.5$\pm$1.0 & 46.7$\pm$0.2 & 1010$\pm$50 & 50.3$\pm$2.4 & 23.0$\pm$1.3 & 750$\pm$40 & 38.6$\pm$1.7 & 17.8$\pm$1.2 & \\
0808$-$078 & & & & 710$\pm$40 & 15.3$\pm$2.0 & 43$\pm$10 & 700$\pm$30 & 23.5$\pm$1.6 & 43.6$\pm$1.1 & \\
0813+482 &1960$\pm$60 & 24.9$\pm$0.5 & 11.9$\pm$0.5 & 440$\pm$40 & 12.0$\pm$1.1 & 134.4$\pm$2.2 & 172$\pm$16 & 6.1$\pm$0.8 & 119$\pm$4 & 22\\
0825+031 &1218$\pm$24 & 84.3$\pm$1.5 & 54.6$\pm$0.3 & 1460$\pm$70 & 78$\pm$4 & 35.6$\pm$0.9 & 1340$\pm$70 & 45.9$\pm$2.6 & 39.8$\pm$0.7 & \\
0831+241 &- & - & - & - & - & - & 1360$\pm$70 & 35.8$\pm$2.4$\dagger$ & 13.2$\pm$2.0 & 23\\
0836$-$202 &2730$\pm$50 & 70.3$\pm$1.0 & 111.3$\pm$0.4 & 1790$\pm$90 & 44$\pm$3$\dagger$ & 23.4$\pm$2.0 & 1020$\pm$50 & 9.1$\pm$1.3$\dagger$ & 161$\pm$4 & 24\\
0838+583 &{\em 518$\pm$8} & {\em 19.1$\pm$0.1} & {\em 179.8$\pm$0.1} & 1160$\pm$60 & 29.3$\pm$2.0 & 167.1$\pm$1.7 & 810$\pm$40 & 18.5$\pm$1.1 & 158.6$\pm$1.7 & \\
0840+132 & & & & 710$\pm$40 & 13.0$\pm$1.3 & 13.3$\pm$2.7 & 670$\pm$30 & 18.1$\pm$1.7 & 166.8$\pm$2.1 & \\
0841+708 &1610$\pm$30 & 63.3$\pm$1.2 & 96.3$\pm$0.2 & 1650$\pm$80 & 55.7$\pm$2.5 & 111.9$\pm$1.3 & 2020$\pm$100 & 52.0$\pm$2.1 & 118.0$\pm$1.1 & \\
0854+201 &{\em 3910$\pm$290} & {\em 98.9$\pm$0.2} & {\em 79.9$\pm$0.0} & 3020$\pm$150 & 96$\pm$5 & 178.1$\pm$0.5 & 2620$\pm$130 & 116$\pm$6 & 178.3$\pm$0.3 & \\
0902$-$142 &1383$\pm$28 & 66.7$\pm$1.6 & 84.7$\pm$0.2 & 581$\pm$30 & 10.2$\pm$1.7$\dagger$ & 57$\pm$4 & 356$\pm$18 & 10.5$\pm$1.7$\dagger$ & 102$\pm$4 & \\
0907$-$203 & & & & 338$\pm$17 & 17.0$\pm$2.6$\dagger$ & 14$\pm$6 & 96$\pm$17 &  $<$10 &  - & 25\\
0909+013 &1295$\pm$26 & 63.9$\pm$1.0 & 112.3$\pm$0.4 & 1400$\pm$70 & 36.2$\pm$2.1 & 139.5$\pm$1.6 & 1260$\pm$60 & 23.8$\pm$1.5 & 153.3$\pm$1.7 & \\
0909+428 & & & & 970$\pm$50 & 25.1$\pm$1.9 & 126.2$\pm$1.6 & 740$\pm$40 & 32.9$\pm$2.0 & 146.3$\pm$1.8 & \\
0914+028 &1194$\pm$24 & 44.7$\pm$0.8 & 104.6$\pm$0.4 & 840$\pm$40 & 25.8$\pm$1.8 & 119.7$\pm$2.5 & 581$\pm$29 & 20.2$\pm$1.5 & 119.7$\pm$2.4 & \\
0918$-$120 &2460$\pm$120 & 121.3$\pm$2.3 & 101.8$\pm$0.4 & 500$\pm$40 & 52$\pm$3 & 125.8$\pm$1.7 & 39$\pm$7 &  $<$10 &  - & 26\\
0920+446 &1373$\pm$27 & 31.6$\pm$0.5 & 162.6$\pm$0.4 & 2180$\pm$110 & 25.4$\pm$2.2 & 126.3$\pm$2.0 & 1960$\pm$100 & 30.1$\pm$1.9 & 93.3$\pm$2.5 & \\
0921$-$263 &2090$\pm$40 & 7.5$\pm$0.3 & 172.5$\pm$1.1 & 1190$\pm$60 & 30$\pm$3 & 78.8$\pm$2.9 & 660$\pm$30 & 14.3$\pm$1.9 & 165$\pm$4 & 27\\
0927+390 &{\em 8456$\pm$13} & {\em 195.7$\pm$0.4} & {\em 130.2$\pm$0.0} & 9000$\pm$450 & 315$\pm$14 & 148.6$\pm$1.0 & 5670$\pm$290 & 212$\pm$9 & 148.4$\pm$1.1 & 28\\
0948+406 &1910$\pm$40 & 90.6$\pm$1.7 & 35.2$\pm$0.3 & 1590$\pm$80 & 48.4$\pm$2.5 & 18.3$\pm$1.5 & 1110$\pm$60 & 29.0$\pm$1.5 & 22.9$\pm$1.5 & \\
0955+695 &1000$\pm$30 & 19.1$\pm$0.5 & 40.2$\pm$0.4 & 270$\pm$20 &  $<$10 &  - & 61$\pm$6 &  $<$10 &  - & 29\\
0958+473 &1498$\pm$30 & 38.6$\pm$0.6 & 19.5$\pm$0.5 & 970$\pm$50 & 10.1$\pm$1.1 & 150.0$\pm$3.0 & 585$\pm$29 & 7.0$\pm$1.1 & 156$\pm$5 & \\
1014$-$231 &- & - & - & - & - & - & - & - & - & 30\\
1037$-$295 & & & & - & - & - & - & - & - & 31\\
1038+051 &{\em 491.8$\pm$0.3} & {\em 4.0$\pm$0.2} & {\em 143.4$\pm$1.4} & 1180$\pm$60 & 99$\pm$7 & 50$\pm$7 & 700$\pm$40 & 72$\pm$4 & 176.4$\pm$0.6 & \\
1041+061 &1530$\pm$30 & 71.6$\pm$1.3 & 80.1$\pm$0.3 & 1150$\pm$60 & 20.4$\pm$1.8 & 36.3$\pm$2.8 & 600$\pm$30 & 12.6$\pm$1.9 & 34.1$\pm$2.7 & \\
1047$-$191 &1166$\pm$23 & 31.4$\pm$0.6 & 75.8$\pm$0.4 & - & - & - & - & - & - & 32\\
1047+717 &1215$\pm$24 & 32.5$\pm$0.7 & 37.3$\pm$0.3 & 1370$\pm$70 & 32.9$\pm$2.1 & 159.9$\pm$1.8 & 1140$\pm$60 & 31.1$\pm$1.9 & 146.5$\pm$1.4 & \\
1058+015 &{\em 3853.1$\pm$0.4} & {\em 46.9$\pm$0.1} & {\em 122.4$\pm$0.1} & 4370$\pm$220 & 347$\pm$16 & 125.2$\pm$0.7 & 3340$\pm$170 & 335$\pm$14 & 123.4$\pm$0.7 & \\
1118$-$125 & & & & - & - & - & - & - & - & 33\\
1127$-$189 &2250$\pm$50 & 69.4$\pm$1.4 & 84.2$\pm$0.3 & 1630$\pm$80 &  $<$10 &  - & 830$\pm$40 &  $<$10 &  - & 34\\
1130$-$148 &2970$\pm$60 & 109.5$\pm$2.4 & 131.6$\pm$0.1 & 1460$\pm$70 & 56$\pm$9 & 167.8$\pm$2.8 & 561$\pm$28 & 29$\pm$4 & 158$\pm$4 & 35\\
1130+382 &{\em 882.5$\pm$0.8} & {\em 2.5$\pm$0.2} & {\em 44.7$\pm$2.7} & 1090$\pm$50 & 23.2$\pm$1.6 & 127.3$\pm$1.7 & 660$\pm$30 & 17.6$\pm$1.2 & 118.6$\pm$2.1 & \\
1153+495 &1064$\pm$23 & 28.3$\pm$0.7 & 90.6$\pm$0.2 & 990$\pm$50 & 10.1$\pm$1.2 & 14$\pm$3 & 650$\pm$30 & 12.6$\pm$1.9 & 26$\pm$5 & \\
1155+810 &1464$\pm$29 & 35.1$\pm$0.6 & 74.0$\pm$0.4 & 930$\pm$50 & 22.2$\pm$2.4 & 86.0$\pm$2.3 & 564$\pm$28 & 17.3$\pm$2.8 & 74$\pm$5 & 36\\
1159+292 &{\em 1232.8$\pm$0.1} & {\em 7.7$\pm$0.1} & {\em 167.3$\pm$0.2} & 2580$\pm$130 & 41$\pm$7 & 153$\pm$6 & 2260$\pm$110 & 64.9$\pm$2.7 & 160.8$\pm$1.1 & \\
1209$-$240 &576$\pm$12 & 16.6$\pm$0.6 & 170.2$\pm$0.7 & 445$\pm$23 &  $<$10 &  - & 373$\pm$19 &  $<$10 &  - & \\
1215$-$174 & & & & 1630$\pm$80 & 61$\pm$5 & 74.1$\pm$2.1 & 1240$\pm$60 & 66$\pm$3 & 60.8$\pm$1.2 & \\
1219+058 &340$\pm$7 & 2.2$\pm$0.6 & 76$\pm$7 & 381$\pm$19 &  $<$10 &  - & 222$\pm$11 &  $<$10 &  - & 37\\
1229+020 &31730$\pm$810 & 1560$\pm$30 & 134.2$\pm$0.1 & 24510$\pm$1230 & 1130$\pm$60 & 137.1$\pm$0.4 & 15280$\pm$760 & 920$\pm$40 & 126.3$\pm$0.6 & 38\\
1230+123 & & & & 3190$\pm$320 & 70$\pm$15 & 2$\pm$4 & 1490$\pm$70 & 46$\pm$5 & 158.9$\pm$3.0 & 39\\
1239+074 &876$\pm$18 & 47.4$\pm$1.0 & 138.4$\pm$0.6 & 690$\pm$30 & 55$\pm$4 & 127.8$\pm$1.1 & 423$\pm$21 & 29.7$\pm$2.5 & 114.8$\pm$2.3 & \\
1246$-$257 &871$\pm$18 & 24.0$\pm$0.4 & 22.1$\pm$0.5 & 830$\pm$40 & 19$\pm$26 & 54$\pm$14 & 760$\pm$40 & 32.7$\pm$2.5 & 43.5$\pm$1.6 & \\
1256$-$057 &13950$\pm$280 & 648$\pm$9 & 70.4$\pm$0.4 & 16800$\pm$850 & 760$\pm$40 & 49.3$\pm$0.6 & - & - & - & 40\\
1258$-$223 & & & & 700$\pm$40 & 15$\pm$13$\dagger$ & 27$\pm$19 & 578$\pm$29 & 11.5$\pm$2.8$\dagger$ & 61$\pm$6 & 41\\
1258$-$319 & & & & 980$\pm$50 & 104$\pm$16 & 70$\pm$5 & 573$\pm$29 & 60$\pm$4 & 56.2$\pm$1.4 & \\
1310+323 &{\em 2277.1$\pm$0.4} & {\em 15.3$\pm$0.1} & {\em 23.4$\pm$0.1} & 2110$\pm$110 & 59.2$\pm$2.9 & 31.6$\pm$1.1 & 2330$\pm$120 & 103$\pm$5 & 41.2$\pm$0.7 & \\
1316$-$336 &1820$\pm$40 & 80.2$\pm$1.5 & 168.1$\pm$0.4 & - & - & - & 1730$\pm$90 & 65$\pm$8 & 118$\pm$5 & 42\\
1330+250 & & & & 960$\pm$50 & 25.8$\pm$1.9 & 0.8$\pm$1.8 & 556$\pm$28 & 14.1$\pm$1.3 & 5.4$\pm$1.8 & \\
1331+305 &5210$\pm$100 & 717$\pm$12 & 33.0$\pm$0.3 & 2520$\pm$130 & 306$\pm$13 & 32.7$\pm$0.8 & 1480$\pm$70 & 194$\pm$8 & 32.6$\pm$0.8 & 43\\
1332+020 &556$\pm$11 & 16.8$\pm$0.4 & 177.7$\pm$0.4 & 800$\pm$40 & 5.5$\pm$1.3$\dagger$ & 162$\pm$6 & 890$\pm$40 & 13.6$\pm$1.5$\dagger$ & 61$\pm$3 & \\
1336$-$339 &90$\pm$3 &  $<$10 &  - & 22$\pm$6 &  $<$10 &  - & 8$\pm$3 &  $<$10 &  - & 44\\
1337$-$129 & & & & 5850$\pm$290 & 164$\pm$8 & 42.5$\pm$0.5 & 5470$\pm$270 & 178$\pm$7 & 29.0$\pm$1.0 & \\
1347+123 & & & & 1130$\pm$60 & 14.4$\pm$1.7 & 113$\pm$3 & 760$\pm$40 & 18.7$\pm$1.4 & 17.3$\pm$1.8 & \\
1354$-$106 &884$\pm$18 & 48.5$\pm$0.8 & 152.7$\pm$0.4 & 750$\pm$40 & 17.3$\pm$2.0 & 36$\pm$4 & 650$\pm$30 & 8.0$\pm$1.6 & 15$\pm$5 & \\
1356+193 &1820$\pm$40 & 54.6$\pm$1.2 & 46.9$\pm$0.2 & 2070$\pm$100 & 32.8$\pm$1.9 & 69.0$\pm$1.6 & 1680$\pm$80 & 38.5$\pm$2.1 & 79.3$\pm$1.0 & \\
1408$-$078 &811$\pm$16 & 42.9$\pm$0.8 & 166.2$\pm$0.4 & 770$\pm$40 & 13.5$\pm$1.5$\dagger$ & 66$\pm$3 & 640$\pm$30 & 18.5$\pm$1.4$\dagger$ & 24.9$\pm$2.1 & 45\\
\end{tabular}
\end{table*}

\clearpage
\begin{table*}
\begin{tabular}{lcccccccccc}
&\multicolumn{3}{c}{X-band}&\multicolumn{3}{c}{K-band}&\multicolumn{3}{c}{Q-band}\\
&I(mJy)&P(mJy)&Angle($^{\circ}$)&I(mJy)&P(mJy)&Angle($^{\circ}$)&I(mJy)&P(mJy)&Angle($^{\circ}$)\\ \hline
1419+383 &283$\pm$6 & 7.2$\pm$0.3 & 46.7$\pm$1.0 & 249$\pm$13 &  $<$10 &  - & 209$\pm$17 &  $<$10 &  - & \\
1427$-$330 &31.7$\pm$1.9 & 1.0$\pm$0.2 & 173$\pm$7 & - & - & - & 4.4$\pm$1.9 &  $<$10 &  - & 46\\
1446$-$163 &574$\pm$11 & 14.5$\pm$0.3 & 153.8$\pm$0.5 & 323$\pm$16 & 12.6$\pm$1.5 & 88$\pm$3 & 196$\pm$10 & 7.6$\pm$1.9 & 62$\pm$9 & \\
1458+716 & & & & 1090$\pm$50 & 31.9$\pm$2.6 & 49.3$\pm$1.4 & 690$\pm$40 & 17.0$\pm$1.5 & 40.5$\pm$2.1 & \\
1504+105 &1540$\pm$30 & 53.7$\pm$0.8 & 155.9$\pm$0.4 & 1750$\pm$90 & 32.3$\pm$1.9 & 105.0$\pm$1.8 & 1700$\pm$90 & 56.7$\pm$2.4 & 72.9$\pm$1.1 & \\
1506$-$167 & & & & 870$\pm$40 &  $<$10 &  - & 650$\pm$30 &  $<$10 &  - & \\
1510$-$057 &1730$\pm$40 & 18.8$\pm$0.3 & 109.9$\pm$0.5 & 1190$\pm$60 & 50.5$\pm$2.8 & 86.7$\pm$0.7 & 840$\pm$40 & 40.7$\pm$2.0 & 78.0$\pm$1.1 & \\
1512$-$090 &2880$\pm$60 & 15.7$\pm$0.4 & 117.4$\pm$0.7 & 3060$\pm$150 & 93$\pm$5 & 112.1$\pm$1.5 & 2560$\pm$130 & 80$\pm$3 & 114.6$\pm$1.1 & \\
1513$-$100 &1121$\pm$22 & 44.9$\pm$2.3 & 157.5$\pm$1.5 & 1220$\pm$60 & 36.2$\pm$2.6 & 154.2$\pm$1.9 & 1080$\pm$50 & 18.8$\pm$1.4 & 154.6$\pm$2.1 & \\
1516+002 &{\em 954.6$\pm$1.0} & {\em 3.9$\pm$0.2} & {\em 106.4$\pm$1.6} & 910$\pm$50 & 18$\pm$3 & 58$\pm$7 & 870$\pm$40 & 17.9$\pm$1.5 & 50.6$\pm$1.9 & \\
1517$-$243 & & & & 2150$\pm$110 & 87$\pm$5 & 46.0$\pm$0.8 & 1940$\pm$100 & 68$\pm$3 & 33.7$\pm$1.0 & \\
1540+147 &1610$\pm$30 & 169.5$\pm$2.5 & 154.8$\pm$0.4 & 1270$\pm$60 & 125$\pm$6 & 142.5$\pm$0.6 & 1000$\pm$50 & 99$\pm$5 & 140.5$\pm$0.5 & \\
1549+026 &{\em 975.2$\pm$0.8} & {\em 5.6$\pm$0.1} & {\em 50.6$\pm$0.5} & 2420$\pm$120 & 67$\pm$7 & 61$\pm$5 & 2170$\pm$110 & 73$\pm$4 & 51.2$\pm$0.6 & \\
1550+054 &3300$\pm$70 & 240$\pm$4 & 144.9$\pm$0.3 & 2710$\pm$140 & 180$\pm$9 & 141.3$\pm$0.5 & 2170$\pm$110 & 139$\pm$7 & 139.8$\pm$0.4 & \\
1608+104 &1497$\pm$30 & 21.7$\pm$0.5 & 161.7$\pm$0.6 & 1260$\pm$60 & 12.7$\pm$1.3$\dagger$ & 74.9$\pm$2.6 & 980$\pm$50 & 12.5$\pm$1.0$\dagger$ & 65.7$\pm$2.4 & 47\\
1613+342 &{\em 3060$\pm$130} & {\em 86.8$\pm$0.4} & {\em 4.8$\pm$0.0} & 2610$\pm$130 & 38.7$\pm$2.4 & 25.2$\pm$1.8 & 1680$\pm$80 & 22.5$\pm$1.6 & 36.3$\pm$2.6 & \\
1633+824 &781$\pm$16 & 13.6$\pm$0.3 & 110.3$\pm$0.6 & 820$\pm$40 &  $<$10 &  - & - & - & - & 48\\
1635+381 &2770$\pm$60 & 96.9$\pm$1.9 & 52.5$\pm$0.2 & 2770$\pm$140 & 23.3$\pm$1.8 & 53.0$\pm$1.5 & 2670$\pm$130 & 56.5$\pm$2.4 & 64.6$\pm$1.2 & \\
1638+573 &{\em 1356$\pm$7} & {\em 27.5$\pm$0.1} & {\em 125.7$\pm$0.1} & 2210$\pm$110 & 12.8$\pm$1.5 & 98$\pm$5 & 1880$\pm$90 & 22.2$\pm$2.4 & 88.5$\pm$1.4 & \\
1642+398 &{\em 5653.8$\pm$1.9} & {\em 238.5$\pm$0.5} & {\em 24.8$\pm$0.1} & 4540$\pm$230 & 79$\pm$4 & 38.9$\pm$0.8 & 4200$\pm$210 & 77$\pm$4 & 48.9$\pm$0.5 & 49\\
1642+689 & & & & 4660$\pm$230 & 54$\pm$4 & 148.0$\pm$2.1 & 4670$\pm$230 & 270$\pm$11 & 103.2$\pm$0.8 & \\
1651+049 &1630$\pm$90 & 195$\pm$5 & 26.7$\pm$0.7 & 231$\pm$18 &  $<$10 &  - & 18.3$\pm$2.6 &  $<$10 &  - & 50\\
1654+396 &1311$\pm$26 & 36.5$\pm$0.6 & 14.3$\pm$0.4 & 1020$\pm$50 & 33.4$\pm$2.2 & 138.0$\pm$1.1 & 860$\pm$40 & 22.9$\pm$1.6 & 140.9$\pm$1.9 & \\
1657+479 &743$\pm$15 & 8.8$\pm$0.2 & 71.5$\pm$0.7 & 670$\pm$30 & 15.1$\pm$0.7 & 148.5$\pm$1.3 & 630$\pm$30 & 12.9$\pm$0.8 & 135$\pm$8 & \\
1658+077 &{\em 917.9$\pm$0.4} & {\em 16.5$\pm$0.1} & {\em 153.0$\pm$0.1} & 1800$\pm$90 & 92$\pm$10 & 113.6$\pm$2.8 & 1590$\pm$80 & 82$\pm$3 & 114.1$\pm$1.1 & \\
1734+389 &955$\pm$19 & 6.1$\pm$0.3 & 3.2$\pm$1.1 & 1230$\pm$60 & 67.7$\pm$2.8 & 107.1$\pm$1.1 & 1220$\pm$60 & 98$\pm$4 & 101.5$\pm$0.8 & \\
1740+522 &{\em 1357.8$\pm$0.6} & {\em 15.9$\pm$0.1} & {\em 21.1$\pm$0.2} & 910$\pm$50 & 15.6$\pm$1.4 & 151.8$\pm$2.5 & 730$\pm$40 & 20.9$\pm$1.3 & 160.3$\pm$1.8 & \\
1753+288 &{\em 518.8$\pm$1.0} & {\em 27.1$\pm$0.1} & {\em 162.1$\pm$0.1} & 1740$\pm$90 & 62$\pm$5 & 156.6$\pm$2.4 & 1380$\pm$70 & 38.6$\pm$1.7 & 157.4$\pm$1.3 & \\
1800+784 &3000$\pm$60 & 118.2$\pm$1.7 & 111.7$\pm$0.4 & 2940$\pm$150 & 115$\pm$5 & 99.9$\pm$0.8 & 2380$\pm$120 & 98$\pm$5 & 97.2$\pm$0.6 & \\
1801+440 &1329$\pm$27 & 62.4$\pm$1.2 & 49.9$\pm$0.2 & 1350$\pm$70 & 44$\pm$9 & 44.2$\pm$0.5 & 1300$\pm$70 & 27.7$\pm$1.9 & 48.5$\pm$2.6 & \\
1806+698 &1720$\pm$40 & 59.3$\pm$1.0 & 151.4$\pm$0.4 & 1620$\pm$80 & 7.5$\pm$1.3 & 147$\pm$4 & 1460$\pm$70 & 8.4$\pm$1.1 & 30$\pm$4 & 51\\
1824+568 &1385$\pm$28 & 50.4$\pm$1.0 & 33.2$\pm$0.3 & 1450$\pm$70 & 111$\pm$4 & 19.4$\pm$1.1 & 1230$\pm$60 & 116$\pm$5 & 15.4$\pm$0.9 & \\
1829+487 &3370$\pm$80 & 71.2$\pm$1.2 & 63.2$\pm$0.4 & 2510$\pm$130 & 50.2$\pm$2.6 & 101.1$\pm$1.1 & 1750$\pm$90 & 57$\pm$3 & 91.8$\pm$0.5 & 52\\
1842+681 &1640$\pm$30 & 88.8$\pm$1.6 & 9.9$\pm$0.3 & 1650$\pm$80 & 100$\pm$4 & 15.9$\pm$1.0 & 1440$\pm$70 & 74$\pm$4 & 8.3$\pm$0.7 & \\
1849+670 &1487$\pm$30 & 34.0$\pm$0.7 & 143.9$\pm$0.4 & 2120$\pm$110 & 10.5$\pm$1.3$\dagger$ & 117$\pm$4 & 2100$\pm$100 & 32.4$\pm$1.5$\dagger$ & 155.4$\pm$1.3 & 53\\
1850+283 &1493$\pm$30 & 31.2$\pm$0.6 & 23.4$\pm$0.5 & 840$\pm$40 &  $<$10 &  - & 432$\pm$22 &  $<$10 &  - & \\
1902+318 &1590$\pm$30 & 50.5$\pm$1.0 & 44.2$\pm$0.2 & 990$\pm$50 & 14.6$\pm$1.2 & 27.9$\pm$2.4 & 660$\pm$30 & 14.0$\pm$1.2 & 43.2$\pm$2.0 & \\
1923$-$210 & & & & 1520$\pm$80 & 31$\pm$4 & 134.1$\pm$1.4 & 1450$\pm$70 & 16$\pm$5 & 9$\pm$10 & \\
1924$-$292 &11920$\pm$240 & 224$\pm$4 & 147.0$\pm$0.3 & 13670$\pm$680 & 1010$\pm$50 & 84.9$\pm$0.4 & 12870$\pm$640 & 1150$\pm$50 & 81.4$\pm$0.6 & \\
1927+613 &1035$\pm$21 & 65.9$\pm$1.2 & 126.6$\pm$0.2 & 810$\pm$40 & 38.6$\pm$2.0 & 112.4$\pm$1.5 & 583$\pm$29 & 27.2$\pm$1.8 & 96.5$\pm$1.6 & \\
1927+739 &3220$\pm$60 & 44.3$\pm$0.7 & 117.8$\pm$0.4 & 2820$\pm$140 & 43.3$\pm$2.2 & 67.4$\pm$1.4 & 2230$\pm$110 & 50.9$\pm$2.6 & 54.5$\pm$1.3 & \\
1939$-$154 &789$\pm$16 & 12.8$\pm$0.2 & 157.0$\pm$0.5 & 720$\pm$40 & 6.8$\pm$1.4$\dagger$ & 62$\pm$6 & 509$\pm$25 & 7.4$\pm$1.1$\dagger$ & 30$\pm$3 & \\
2000$-$178 &2040$\pm$40 & 72.9$\pm$1.2 & 148.8$\pm$0.4 & 2010$\pm$100 & 15.9$\pm$2.1$\dagger$ & 62$\pm$4 & 1790$\pm$90 & 26.8$\pm$1.7$\dagger$ & 18.3$\pm$1.8 & 54\\
2011$-$157 &2080$\pm$40 & 46.0$\pm$1.1 & 149.7$\pm$0.9 & 1810$\pm$90 & 37$\pm$10 & 125$\pm$3 & 1360$\pm$70 & 22.1$\pm$2.9 & 95$\pm$8 & \\
2022+616 &{\em 3028.5$\pm$1.2} & {\em 0.0$\pm$0.1} & {\em 180$\pm$70} & 1810$\pm$90 & 6.9$\pm$1.3 & 59$\pm$5 & 1000$\pm$50 & 6.5$\pm$1.1 & 51$\pm$5 & \\
2101+037 &831$\pm$17 & 41.0$\pm$0.8 & 130.7$\pm$0.2 & 720$\pm$40 & 33.6$\pm$1.8 & 115.5$\pm$1.5 & 620$\pm$30 & 13.5$\pm$1.3 & 102.3$\pm$2.6 & \\
2123+056 &1910$\pm$40 & 56.6$\pm$1.2 & 5.2$\pm$0.2 & 1460$\pm$70 & 82$\pm$3 & 28.1$\pm$1.0 & 1120$\pm$60 & 69.2$\pm$2.6 & 21.0$\pm$1.1 & \\
2131$-$121 & & & & 1980$\pm$100 & 41.5$\pm$2.9 & 149.4$\pm$1.6 & 1600$\pm$80 & 35.9$\pm$2.1 & 174.3$\pm$1.2 & \\
2134$-$019 & & & & 2040$\pm$100 & 162$\pm$8 & 91.0$\pm$0.2 & 1620$\pm$80 & 130$\pm$5 & 72.4$\pm$1.0 & \\
2136+006 &{\em 7549$\pm$12} & {\em 44.7$\pm$0.4} & {\em 161.7$\pm$0.2} & 5220$\pm$260 & 134$\pm$5 & 18.0$\pm$1.0 & 3180$\pm$160 & 61$\pm$3 & 0.8$\pm$0.4 & \\
2139+144 &{\em 2515.8$\pm$0.7} & {\em 14.2$\pm$0.1} & {\em 32.1$\pm$0.2} & 2110$\pm$110 & 38$\pm$9 & 19$\pm$5 & 1310$\pm$70 & 40.8$\pm$2.2 & 172.8$\pm$1.0 & \\
2143+176 &561$\pm$11 & 9.7$\pm$0.5 & 142.4$\pm$0.8 & 690$\pm$30 & 24.3$\pm$2.0 & 11.4$\pm$1.7 & 620$\pm$30 & 17.4$\pm$1.9 & 13.7$\pm$2.1 & \\
2148+069 &6220$\pm$120 & 122.2$\pm$2.3 & 126.2$\pm$0.3 & 5250$\pm$260 & 95$\pm$4 & 66.3$\pm$1.2 & 4690$\pm$230 & 54.5$\pm$2.9 & 41.9$\pm$1.4 & \\
2151$-$304 &1750$\pm$40 & 39.4$\pm$0.6 & 67.5$\pm$0.4 & 1740$\pm$90 & 35$\pm$4 & 30.7$\pm$2.6 & 1300$\pm$70 & 33.0$\pm$2.4 & 17.0$\pm$1.9 & \\
2158$-$150 &1970$\pm$40 & 71.2$\pm$1.8 & 71.6$\pm$0.7 & 1360$\pm$70 & 62$\pm$3 & 44.7$\pm$0.6 & 1120$\pm$60 & 61.6$\pm$2.4 & 23.3$\pm$1.1 & \\
2202+422 & & & & 2690$\pm$130 & 184$\pm$7 & 17.8$\pm$1.0 & 2410$\pm$120 & 180$\pm$8 & 12.5$\pm$0.8 & \\
2203+173 &1346$\pm$27 & 44.7$\pm$0.9 & 7.3$\pm$0.2 & 1340$\pm$70 & 67.4$\pm$2.6 & 20.3$\pm$1.1 & 1090$\pm$50 & 70.5$\pm$3.0 & 14.1$\pm$0.9 & \\
2203+317 &2700$\pm$50 & 46.4$\pm$0.9 & 37.8$\pm$0.2 & 2140$\pm$110 & 19.1$\pm$1.6 & 87.5$\pm$1.4 & 1320$\pm$70 & 44.0$\pm$1.9 & 110.4$\pm$1.2 & \\
2206$-$186 &3150$\pm$60 & 41.6$\pm$0.9 & 55.0$\pm$0.7 & 1720$\pm$90 & 60$\pm$4 & 40.5$\pm$0.9 & 1080$\pm$50 & 33.1$\pm$2.4 & 32.2$\pm$1.5 & \\
2211+238 &1220$\pm$24 & 66.9$\pm$1.3 & 6.5$\pm$0.2 & 950$\pm$50 & 47.3$\pm$2.0 & 27.1$\pm$1.2 & 640$\pm$30 & 36.1$\pm$1.7 & 24.0$\pm$1.3 & \\
2218$-$035 & & & & 1640$\pm$80 & 15$\pm$3 & 75$\pm$9 & 1180$\pm$60 & 6.9$\pm$1.4 & 49$\pm$5 & \\
2225$-$049 &7510$\pm$150 & 161.3$\pm$2.8 & 163.2$\pm$0.4 & 8630$\pm$430 & 264$\pm$10 & 19.6$\pm$1.0 & 7620$\pm$380 & 204$\pm$8 & 17.6$\pm$1.0 & \\
\end{tabular}
\end{table*}

\clearpage
\begin{table*}
\begin{tabular}{lcccccccccc}
&\multicolumn{3}{c}{X-band}&\multicolumn{3}{c}{K-band}&\multicolumn{3}{c}{Q-band}\\
&I(mJy)&P(mJy)&Angle($^{\circ}$)&I(mJy)&P(mJy)&Angle($^{\circ}$)&I(mJy)&P(mJy)&Angle($^{\circ}$)\\ \hline
2229$-$085 &3390$\pm$70 & 107.5$\pm$2.1 & 139.1$\pm$0.1 & 3040$\pm$150 & 55$\pm$3 & 178.1$\pm$0.8 & 3210$\pm$160 & 62$\pm$3 & 7.3$\pm$0.8 & \\
2232+117 &{\em 3029.9$\pm$2.4} & {\em 39.1$\pm$0.5} & {\em 89.0$\pm$0.2} & 4860$\pm$240 & 88$\pm$4 & 67.9$\pm$1.3 & 4300$\pm$220 & 85$\pm$7 & 92.8$\pm$0.6 & \\
2236+284 &{\em 2160.5$\pm$0.2} & {\em 33.7$\pm$0.1} & {\em 51.8$\pm$0.1} & 1330$\pm$70 & 25.7$\pm$2.0 & 163.9$\pm$2.0 & 1140$\pm$60 & 52.0$\pm$2.9 & 175.2$\pm$1.0 & \\
2246$-$121 &2420$\pm$50 & 27.3$\pm$0.7 & 119.9$\pm$0.5 & 2350$\pm$120 & 30.3$\pm$2.0 & 2.8$\pm$2.3 & 1980$\pm$100 & 61.5$\pm$2.5 & 163.7$\pm$1.0 & \\
2254+161 &9660$\pm$190 & 470$\pm$9 & 176.4$\pm$0.1 & 15520$\pm$780 & 269$\pm$19 & 140.9$\pm$0.6 & 20620$\pm$1030 & 250$\pm$30 & 101$\pm$3 & 55\\
2255+420 &727$\pm$15 & 15.7$\pm$0.4 & 105.8$\pm$0.9 & 438$\pm$22 &  $<$10 &  - & 183$\pm$9 &  $<$10 &  - & \\
2258$-$279 & & & & 2860$\pm$140 & 18$\pm$6 & 160$\pm$8 & 3060$\pm$160 & 77$\pm$4 & 166.4$\pm$1.4 & 56\\
2330+109 &{\em 1018.4$\pm$0.1} & {\em 16.4$\pm$0.1} & {\em 38.6$\pm$0.1} & 770$\pm$40 & 23.4$\pm$1.8 & 45.5$\pm$1.0 & 590$\pm$30 & 4.9$\pm$1.3 & 28$\pm$7 & \\
2331$-$159 & & & & 690$\pm$30 & 50$\pm$3 & 72.6$\pm$1.8 & 524$\pm$26 & 32.3$\pm$1.7 & 69.4$\pm$1.6 & \\
2334+075 &1137$\pm$23 & 41.2$\pm$0.8 & 146.2$\pm$0.3 & 1490$\pm$70 & 42.8$\pm$2.7 & 141.4$\pm$0.8 & 1320$\pm$70 & 54.9$\pm$2.5 & 149.2$\pm$1.1 & \\
2346+094 &1281$\pm$26 & 45.2$\pm$0.9 & 130.5$\pm$0.2 & 980$\pm$50 & 30.4$\pm$2.1 & 150.3$\pm$1.7 & 640$\pm$30 & 21.7$\pm$1.3 & 154.0$\pm$1.7 & \\
2348$-$165 &1590$\pm$30 & 10.6$\pm$0.3 & 76.7$\pm$0.8 & 1990$\pm$100 & 28.1$\pm$2.3 & 16.4$\pm$2.5 & 2230$\pm$110 & 51.3$\pm$2.2 & 158.8$\pm$1.2 & \\
2354+458 &1253$\pm$25 & 17.0$\pm$0.5 & 176.9$\pm$0.3 & 770$\pm$40 & 10.1$\pm$1.5 & 101$\pm$4 & 392$\pm$20 &  $<$10 &  - & \\
2358$-$102 &1202$\pm$24 & 31.0$\pm$0.6 & 122.8$\pm$0.3 & 980$\pm$50 & 15.7$\pm$3.0 & 32$\pm$5 & 820$\pm$40 & 21.7$\pm$2.2 & 17$\pm$4 & \\
\end{tabular}
\end{table*}

\clearpage

{\bf Caption to Table 2.}
\footnotesize

Flux densities of observed sources at X-, K- and Q-bands. Fluxes in
italics are from CLASS, in the case of multiple observations using that
with the lowest error in polarized flux density. A dagger indicates that
the measurement has been taken from the $Q$ and $U$ maps, rather than
the fits to the u-v data (see text for details). A ``zero'' error means
smaller than 1~mJy. Notes to the table:
1.  Nearest plausible ID is 4C+25.01 (00 19 39.21 +26 02 45.4) which is 0.5~Jy
at 1.4~GHz (NVSS); multiple sources may contribute to the WMAP flux density. The NVSS
position was observed, leading to the detection of a resolved, polarized source at X-band.
 2.  Source is identified with 3C20 which is a 50$^{\prime\prime}$ 
steep-spectrum double (Laing 1981). Nearly all structure is resolved out by
these observations.
 3.  Identification is the star-forming galaxy NGC253. Structure is nearly 
resolved out by these observations.
 4.  No X-band polarization available from CLASS.
 5.  Identification is 3C33 which is a 5$^{\prime}$ steep spectrum double
(e.g. Leahy \& Perley 1991). Only the central regions are seen in these observations.
 6.  Used as the zero-polarization calibrator at X-band.
 7.  Offset polarized component? Fit to the K-band images gives 13.4~mJy polarized flux
density at 11$^{\circ}$. No significant difference between methods at Q-band.
 8.  CLASS X-band polarization measured on two occasions: 21~mJy in PA 57 (total
flux density 1039~mJy), 29mJy in PA 25 (total flux density 1061~mJy).
 9.  Identification is the pair of sources 3C~66A, 3C~66B. Observation pointed
at 3C~66A and structure heavily resolved; most of the high-frequency contribution to
the flux is probably from 3C~66B, however.
 10.  No X-band polarization available. K-band polarization from fitting to $u-v$ data
probably underestimated; images give 10.2~mJy in PA 25$^{\circ}$.
 11.  Source is identified with 3C78 (e.g. Saikia et al. 1986); heavily resolved
by these observations.
 12.  Identified with 3C84; used as zero-polarization calibrator for K-band and Q-band observations.
Polarization level difficult to measure because of residual errors; best guess from analysis of
images is 13~mJy polarized flux density in PA 129$^{\circ}$ 
 13.  Identified with 3C98 which is a 216 arcsec steep spectrum double (e.g. Leahy et al. 1997);
heavily resolved by these observations.
 14.  Probably resolved; peak in the Q-band image has a polarized flux density of 11mJy at PA 0$^{\circ}$.
 15.  Extreme discrepancy between flux observed in all VLA bands and the WMAP flux. No obvious 
identification from NVSS; the observed position (04 24 2.58 +02 26 42) is the closest NVSS source
(about 10$^{\prime}$) from the WMAP position; WMAP flux may be a combination of this and two other 
$\sim$1-Jy sources at only slightly larger distance.
 16.  Fits to K-band image suggest 12mJy polarized flux density in PA 117$^{\circ}$, but sparse data 
make estimation difficult.
 17.  Blended source (see note in Wright et al. 2008).
 18.  ID is a planetary nebula (IC0418).
 19.  Several X-band measurements from CLASS with highly variable polarized flux density and PA.
 20.  Identification is a very large (15$^{\prime}$) double-lobed radio source, 
PKS0634$-$20. One of the lobes was observed, and is heavily resolved.
 21.  Possible offset polarized component in K-band; analysis of image suggests 10mJy in PA 14$^{\circ}$.
 22.  ID is probably the strong source 4.5$^{\prime}$ SSE, 3C196. This is a small double (e.g. Reid et al
1995) both of whose components are detected.
 23.  Very little data obtained at X-band and K-band, no reliable results at these frequencies.
 24.  Phase errors limit reliability; analysis of images suggests K-band polarization 44mJy at
PA 23$^{\circ}$. No significant polarization detected at Q-band.
 25.  Extremely steep spectral index between K-band and Q-band. 
 26.  ID is Hydra A, a large radio source (e.g. Taylor et al. 1990). Core of the source was observed,
consisting of extended feature 45$^{\prime\prime}$ across. Polarization level depends strongly on
aperture used.
 27.  Analysis of images suggests higher K-band polarization (50~mJy) at same PA.
 28.  Identification is 4C39.25. X-band polarization (CLASS) highly variable.
 29.  Identification is M82 (starburst galaxy).
 30.  Error in pointing of array.
 31.  Clear ID, but data of poor quality, phase solutions not good. Polarization measurements 
unreliable at K and Q-band.
 32.  Poor quality data at K-band and Q-band, phase solutions not good. Polarization measurements
unreliable at both frequencies.
 33.  Most plausible ID is 4$^{\prime}$ W, little data at K-band and phase solutions not good.
 34.  Good ID, poor quality data at K-band and Q-band, phase solutions not good and polarization
measurements unreliable.
 35.  Good ID, poor quality data at Q-band, phase solutions not good and polarization measurements
unreliable.
 36.  Analysis of Q-band image suggest slightly lower polarized flux density, 12mJy at PA 80$^{\circ}$.
 37.  ID is 3C270; pointing position is that of the core (Birkinshaw \& Davies 1985).
 38.  ID is 3C273; polarization at X-band (CLASS) is high and variable.
 39.  ID is M87; the polarized jet pointing WNW is visible. Very little data at K-band; this
polarization measurement is unreliable.
 40.  Sparse data at K and Q-band; image at Q-band suggests that this polarization measurement
is unreliable.
 41.  Fits to Q-band images suggest polarized flux density of 12mJy in PA 61$^{\circ}$.
 42.  Little data at K-band; unreliable polarization measurement at this frequency.
 43.  ID is 3C286; primary flux and polarization position angle calibrator; also used
for non-closing offset calibration at X-band.
 44.  IC4296. This is ~2.5 arcmin double with compact core. No significant polarization detected.
 45.  Possible offset polarized component in K-band; analysis of image suggests 14mJy at PA 66$^{\circ}$.
 46.  No obvious catalogued source. Nearest radio source is MRC~1424$-$328 
(NVSS position 14 27 35.7, $-$33 02 22, 253~mJy at 1.4~GHz) approximately 1.3 
arcminutes E. This position yields 180~mJy of correlated flux at 8.4~GHz and
no detections at higher frequencies. WMAP flux density of 1~Jy at 22~GHz may be
the sum of a number of nearby sources.
 47.  Analysis of K-band image suggests slightly higher polarization: 13mJy at PA 75$^{\circ}$; 
detection is significant.
 48.  ID is NGC6251; position of core observed (e.g. Perley, Bridle \& Willis 1984).
 49.  ID is 3C345. Phase errors in K-band observation; analysis of image suggests slightly
higher polarized flux density of 75mJy in PA 164$^{\circ}$.
 50.  ID is Hercules A; core observed (e.g. Gizani \& Leahy 2003), highly resolved. The
core lies in between two polarized components, and its polarized flux density is much less than
that inferred from the $u-v$ data by the automatic fitting.
 51.  Fits to Q-band images suggest polarized flux density 14mJy in PA 60$^{\circ}$.
 52.  ID is 3C380.
 53.  K-band images reveal significant detection of polarized flux density, $\sim$10mJy in PA 117$^{\circ}$.
 54.  Possibly offset polarized component at K-band of 16mJy in PA 62$^{\circ}$.
 55.  ID is 3C454.3.
 56.  K-band polarization measurement unreliable; noisy map with significant phase errors.

\end{document}